\documentclass[twocolumn, switch]{article} % Method A for two-column formatting
\usepackage{preprint}
\usepackage{algorithm}
\usepackage{algorithmic}
\usepackage{amsmath, amsthm, amssymb, amsfonts}
\usepackage{gensymb}
\usepackage[numbers,square]{natbib}
\usepackage[utf8]{inputenc}	% allow utf-8 input
\usepackage[T1]{fontenc}	% use 8-bit T1 fonts
\usepackage{xcolor}		% colors for hyperlinks
\usepackage[colorlinks = true,
            linkcolor = purple,
            urlcolor  = blue,
            citecolor = cyan,
            anchorcolor = black]{hyperref}	% Color links to references, figures, etc.
\usepackage{booktabs} 		% professional-quality tables
\usepackage{nicefrac}		% compact symbols for 1/2, etc.
\usepackage{microtype}		% microtypography
\usepackage{lineno}		% Line numbers
\usepackage{float}			% Allows for figures within multicol

\usepackage{lipsum}		%  Filler text

\usepackage{newfloat}
\DeclareFloatingEnvironment[name={Supplementary Figure}]{suppfigure}
\usepackage{sidecap}
\sidecaptionvpos{figure}{c}

\usepackage{titlesec}
\titlespacing\section{0pt}{12pt plus 3pt minus 3pt}{1pt plus 1pt minus 1pt}
\titlespacing\subsection{0pt}{10pt plus 3pt minus 3pt}{1pt plus 1pt minus 1pt}
\titlespacing\subsubsection{0pt}{8pt plus 3pt minus 3pt}{1pt plus 1pt minus 1pt}

\title{Arctic bimodality reflects enhanced variability in the clear-cloudy transition rather than bistability}

\usepackage{eso-pic}
\usepackage{tikz}
\usepackage{xcolor}
\PassOptionsToPackage{colorlinks=true,linkcolor=gray,urlcolor=gray}{hyperref}

\usepackage{titling}
\usepackage{orcidlink}
\usepackage{footmisc}
\newcommand{\Author}[3]{% Name, ORCID, Institution
  \textbf{#1}\textsuperscript{#2},\ \orcidlink{#3} %
}

\author{
  \Author{Nadja Omanovic}{1}{0000-0002-6252-5131}\and
  \Author{J. Antoon van Hooft}{2}{0000-0002-3952-0172}\and
  \Author{Valentina Zeni}{3}{0009-0000-7992-1053}\and
  \Author{Franziska Glassmeier}{1,3}{0000-0002-1132-7821}
}

\date{%
  \textsuperscript{1}Max Planck Institute for Meteorology, Hamburg, Germany\\
  \textsuperscript{2}Department of Biomedical Engineering, Eindhoven University of Technology, Eindhoven, the Netherlands\\
  \textsuperscript{3}Department of Geoscience \& Remote Sensing, Delft University of Technology, Delft, the Netherlands\\[1em]
  \footnotesize \textbf{Corresponding author:} Nadja Omanovic \texttt{<nadja.omanovic@mpimet.mpg.de>}\\
}

\begin{document}

\twocolumn[ % Method A for two-column formatting
  \begin{@twocolumnfalse} % Method A for two-column formatting

\maketitle
\thispagestyle{empty}

\begin{abstract}
The wintertime Arctic boundary layer is characterized by two distinct states, radiatively clear (cloud-free) and radiatively opaque (cloudy), which is reflected in observations as a bimodal distribution. Previous studies hypothesized that these regimes are related to an underlying bistability of the system's dynamics. We apply a drift--diffusion framework based on the two leading terms of the Kramers-Moyal expansion to observations from 15 winter periods in Ny-Ålesund, Norway. Our model captures the bimodality; we find a single stable point for the deterministic drift, however, which questions the existence of bistable deterministic dynamics. Instead, the observed regimes can be attributed to state-dependent stochastic fluctuations with maximum amplitude close to the clear-cloudy transition. The position of the stable point depends on the meteorological conditions. Controlling for synoptic variability does not remove the state-dependence of diffusion, however. The cloudy regime is associated with southwesterly winds, high specific humidities, and uplifting motion, the latter two being conducive for cloud formation, while the cloud-free regime occurs under drier conditions and northeasterly winds. Through an illustrative example, we suggest a possible process-based interpretation of our statistical analysis. The nonlinear onset of cloud formation, once atmospheric humidity reaches saturation, could map state-independent fluctuations in humidity to state-dependent fluctuations in cloudiness.
\end{abstract}
%\keywords{First keyword \and Second keyword \and More} % (optional)
\vspace{0.35cm}

\end{@twocolumnfalse} % Method A for two-column formatting
] % Method A for two-column formatting

%\begin{multicols}{2} % Method B for two-column formatting (doesn't play well with line numbers), comment out if using method A

%%%%%%%%%%%%%%%  Main text   %%%%%%%%%%%%%%%
% \linenumbers

% \section{Template Description}

% This LaTeX template is a customized and optimized two-column format for arXiv preprint submissions.

% It is ideal for researchers seeking a reliable and well-tested two-column arXiv template with tailored improvements. The template maintains the clean and professional layout of the original while introducing enhancements for better compatibility, streamlined dependencies, and personalized formatting adjustments.

% To ensure proper bibliography compilation and facilitate submission to platforms like arXiv, it is necessary to obtain the generated \texttt{.bbl} file from your Overleaf project. After successful compilation, open the “Logs and output files” panel on the left side of the Overleaf interface, locate and open the \texttt{.bbl} file to view its contents. Copy the content into a local text file and save it with the \texttt{.bbl} extension, then upload this \texttt{.bbl} file back to your Overleaf project. When downloading the entire project as a zip archive, the \texttt{.bbl} file will be included, facilitating packaging and offline compilation. This process effectively prevents compilation errors caused by missing \texttt{.bbl} files during submission and ensures consistent and correct bibliography formatting.

% \vspace{1em}
% \noindent\textbf{Version 1.0 --- July 10, 2025}

\section{Introduction}  %% \introduction[modified heading if necessary]
The Arctic region warms nearly four times faster than the rest of the globe, a phenomenon known as Arctic amplification \citep{rantanenArcticHasWarmed2022}. Recent evaluations of CMIP6 revealed, however, that most of the models fail to capture this rapid warming \citep{chylekAnnualMeanArctic2022,duffeyRepresentationArcticWinter2025}. Arctic boundary-layer clouds are identified as a key uncertainty because they modulate incoming and outgoing radiation. During Arctic winters (no incoming shortwave radiation), these clouds have a warming effect as they absorb and re-emit longwave radiation \citep{dong10YearClimatology2010}. 

Arctic boundary-layer clouds form under a temperature inversion and have a liquid layer consisting of cloud droplets at cloud top. In the case of mixed-phase clouds, the liquid layer is supercooled and ice crystals form, grow, and sediment out of the cloud as precipitation. The co-existence of cloud droplets and ice crystals makes these clouds thermodynamically unstable because ice has a lower saturation water vapor pressure than liquid water. This difference gives rise to the Wegener--Bergeron--Findeisen (WBF) process \citep{Wegener1911,Bergeron1935,Findeisen1938}, which describes the vapor depositional growth of ice crystals at the expense of evaporating cloud droplets. The WBF process is an efficient pathway to initiate precipitation  \citep{mulmenstadtFrequencyOccurrenceRain2015,heymsfieldContributionsLiquidIce2020} and has the potential to fully glaciate a cloud. The speed of the WBF process depends on temperature, pressure, supersaturation with respect to ice, cloud droplet and ice crystal number concentrations and radii \citep{pinskyAnalyticalInvestigationGlaciation2014}. In theory, the WBF process seems simple to understand, however in the complex environment of clouds, its ability to glaciate a cloud is impeded by a series of interactions across different scales (micro-, meso-, and large-scale). The liquid layer emits longwave radiation, which cools the cloud top causing turbulent mixing, which in turn produces buoyancy. The buoyancy is associated with local updrafts, which generate supersaturated conditions suitable for cloud droplet formation and growth, counteracting the glaciation via the WBF process as the liquid layer is maintained. In the Arctic, humidity inversions above cloud top can be found, such that the entrainment of air (through turbulent mixing) from the free troposphere acts as an additional moisture source for the cloud droplets and ice crystals. Overall, a self-sustaining feedback loop within the liquid layer of Arctic boundary-layer clouds is established \citep{morrisonResiliencePersistentArctic2012}. Thus, the challenges of numerical weather and climate models include \citep{morrisonResiliencePersistentArctic2012,pithanSelectStrengthsBiases2016,shawUsingSatelliteObservations2022}, but are not limited to, accurately describing cloud droplet and ice crystal concentrations and size distributions \citep{ovchinnikovIntercomparisonLargeeddySimulations2014,schulteImprovingArcticMixedphase} and their dependence on aerosols \citep{mauritsenArcticCCNlimitedCloudaerosol2011,stevensModelIntercomparisonCCNlimited2018,pasquierNyAlesundAerosolCloud2022}, resolving the atmospheric boundary layer \citep{duffeyRepresentationArcticWinter2025}, and correctly capturing surface fluxes above open ocean and sea ice covered surfaces \citep{liResponseSimulatedArctic2017}. These examples showcase the challenge in understanding Arctic boundary-layer clouds. They not only require correct representations on the microscale (aerosols, hydrometeor formation), but also on the mesoscale (boundary layer dynamics) and the large-scale (synoptic forcing) as well as for the feedbacks among the scales.   

A possible result of these complex interactions across all three scales was found by
\citet{morrisonResiliencePersistentArctic2012}. They report an `on/off' behavior in Arctic boundary-layer clouds expressed through two distinct radiative states, as measured during the ``Surface Heat Budget of the Arctic Ocean" (SHEBA) experiment, giving rise to a bimodal distribution \citep{stramlerSynopticallyDrivenArctic2011}. One state is characterized by radiatively clear conditions (cloud-free or thin clouds) with longwave emission of ca. $-$45\,$\mathrm{W m^{-2}}$ and the other one by opaque, cloudy conditions and a longwave emission of about $-$5\,$\mathrm{W m^{-2}}$ (see Fig.~4 in \citet{morrisonResiliencePersistentArctic2012}). 
Such a bimodality can indeed emerge from complex interactions and feedbacks across scales in cloud systems \citep[e.g.,][]{bakerBistabilityCCNConcentrations1990a,feingoldReversibilityTransitionsClosed2015,hernandezAerosolMemoryStratocumulus2026}. 

\begin{figure*}[hbt!]
  \centering
  \includegraphics[width=.8\linewidth]{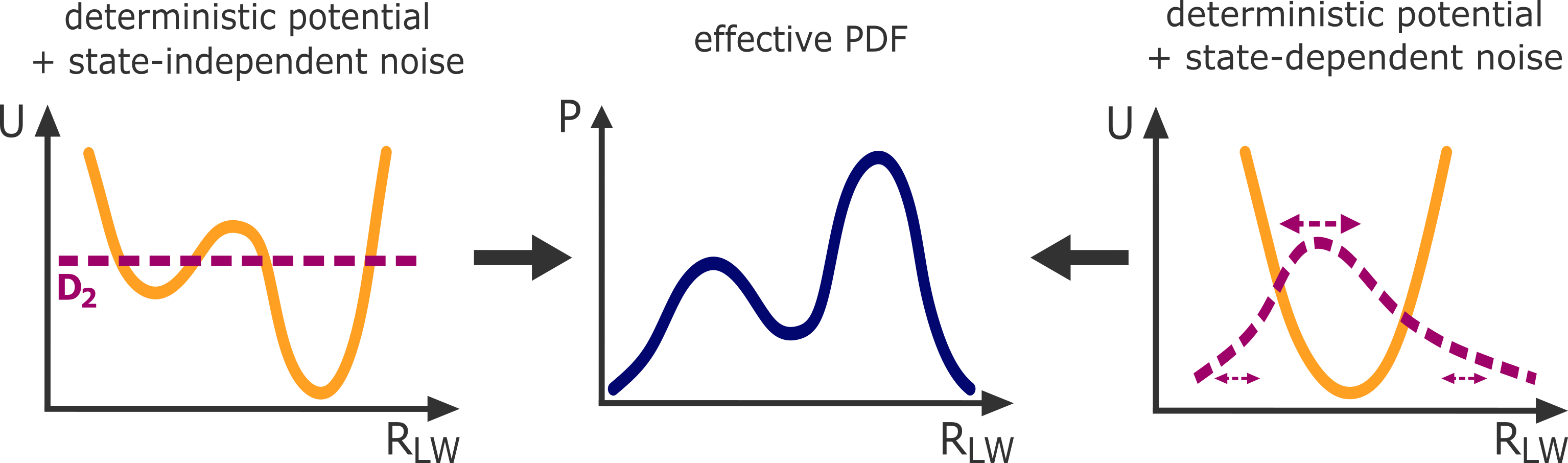}
  \caption{Schematic illustration of the dynamical difference between deterministically (left) and stochastically (right) induced bimodality. Both, a deterministic double-well potential with state-independent noise and a unimodal deterministic potential with state-dependent noise (diffusion coefficient $D_2$) result in a bimodal PDF (middle). R$_\mathrm{{LW}}$ is the radiation ratio (Eq.~\ref{eq:ratio}). Horizontal arrows in the right panel indicate the strength of the noise. Adapted from \citet{suraMultiplicativeNoiseNonGaussianity2005}.}
  \label{fig:sketch}
\end{figure*}

We investigate these complex dynamics utilizing concepts from statistical physics, which also underlie stochastic parameterizations in weather and climate models \citep{bernerStochasticParameterizationNew2017}. Statistical physics explains that bimodality can be either driven by a deterministic double-well potential or by stochastic dynamics in a single-well potential with state-dependent noise \citep{suraMultiplicativeNoiseNonGaussianity2005}, as illustrated in Fig.~\ref{fig:sketch}. Whether the bimodality stems from the one or the other, can be investigated within a drift--diffusion framework that decomposes the system dynamics into its deterministic (drift) and stochastic (diffusion) components.    
In this study, we apply the Langevin equation, a drift--diffusion model, to identify potential stable states of the system directly from observational data via dimensionality reduction \citep[e.g.,][]{araniExitTimeMeasure2021,riechersStableStadialInterstadial2023}. In this approach, the complex dynamics of Arctic boundary-layer clouds are projected onto the ratio of longwave radiation and analyzed as an effective, one-dimensional Langevin process. 

This work is structured as follows. Section \ref{sec:data} introduces the data used in the analysis. In Sects.~\ref{sec:dd_frame} and \ref{sec:ddLE}, an introduction to the drift--diffusion framework is given. The results from its application to the long-term observational data are presented in Sect.~\ref{sec:netRad}, followed by an evaluation of the stochastic model with observations (Sect.~\ref{sec:stoch_model}). Section \ref{sec:syn} presents the Langevin analysis of the observations stratified by meteorological conditions. We demonstrate how to relate our statistical analysis to cloud processes with the help of an illustrative example in Sect.~\ref{sec:toymodel}. Sect.~\ref{sec:concl} discusses and concludes our work.

\section{Data and methodology}
\subsection{Data description} \label{sec:data}
Our analysis focuses on the dynamical behavior of the Arctic cloud system observed in Ny-Ålesund, Svalbard, Norway (78.92$\degree$ N, 11.92$\degree$ E). We use the atmospheric longwave radiation ratio
\begin{equation}\label{eq:ratio}
  \mathrm{R}_\mathrm{{LW}} = \frac{Q_\mathrm{{LWD}}}{Q_\mathrm{{LWU}}}, 
\end{equation}
as a proxy for whether the Arctic boundary layer is cloud-free or cloudy, where Q$_\mathrm{{LWD}}$ and Q$_\mathrm{{LWU}}$ are the downward and upward longwave flux, respectively.

The radiation observations are provided by the Global Energy and Water Cycle Experiment (GEWEX) consortium via Baseline Surface Radiation Network \citep[BSRN,][]{driemel2018baseline}. We focus on the data for the Arctic winter periods (November, December, January, and February) between 2006 and 2021. The observations are normalized by subtracting their global mean and dividing by standard deviation before being discretized into 100 bins, with each bin containing a minimum of 100 data points. We analyze the observations at a time resolution of 10\,min, which is sparsened from the original temporal resolution of 1\,min. The decision for the resampling is based on the Markovian assumption, i.e., the absence of memory in the time series, of the Langevin equation (see Appendix \ref{sec:app}).

The long-term observations allow us to conduct a statistical analysis of the effect of synoptic forcing on the radiative state of the Arctic boundary layer. For this, we retrieve data from the ERA5 reanalysis database at an hourly resolution \citep{hersbach2020era5}. We choose the 900\,hPa level data to represent the synoptic conditions most relevant to dictate the boundary-layer forcing. We analyze the horizontal wind components, the vertical wind component, and the specific humidity, which are normalized by subtracting the mean and dividing by the standard deviation. For this part of the analysis, we interpolate the ERA5 data to 10\,min frequency to match the observations.  

\subsection{The drift--diffusion framework}\label{sec:dd_frame}
Our analysis of the observational data is based on the one-dimensional Langevin equation \citep[LE, e.g.,][]{riskenFokkerPlanckEquationMethods1996}, which is the Itô stochastic differential equation describing the change (increment) of a variable $\phi$ during a small time step $\mathrm{d}t$ at some time $t$ (i.e.,
$\mathrm{d} \phi = \phi (t + \mathrm{d}t) - \phi(t)$). The LE reads
\begin{equation}\label{eq:LE}
  \mathrm{d}\phi = D_1(\phi)\mathrm{d}t + \sqrt{2D_2(\phi)} \mathrm{d}W,  
\end{equation}
where $D_1(\phi)$ is the drift coefficient, describing the average tendency of $\phi$ to increase or decrease as a function of its current state, and $D_2(\phi)$ is the diffusion coefficient, describing the strength of fluctuations around this average evolution. The term $\mathrm{d}W$ corresponds to the increment of a Wiener process, represented by a random sample from a normal distribution. Together, the drift and diffusion terms describe the deterministic and stochastic components of the dynamics, respectively. 

Apart from analyzing a single realization of $\phi(t)$, one can look at an ensemble of such trajectories and consider the probability density to find the system in a certain state $\rho(\phi)$. Its evolution is described by the Fokker-Planck equation \citep[FPE,][]{riskenFokkerPlanckEquationMethods1996}:
\begin{equation}\label{eq:fpe}
  \frac{\partial \rho}{\partial t} = -\frac{\partial \left(D_1\rho\right)}{\partial \phi} + \frac{\partial^2 \left(D_2 \rho\right)}{\partial \phi ^2}.\\
\end{equation}
This statistical framework characterizes the behavior of a large ensemble of stochastic realizations. Assuming a vanishing steady-state probability current, an equilibrated ensemble is described by the steady-state solution of FPE:
\begin{equation}\label{eq:fpe_s}
    \rho_{s}(\phi)=\frac{N_0}{D_2(\phi)}\exp\left(\int^\phi\frac{D_1(\phi')}{D_2(\phi')}d\phi'\right),
\end{equation}
where $N_0$ ensures that $\rho_{s}$ is normalized. This solution holds for both reflecting (finite domain) and natural (infinite domain) boundary conditions, with their only effect being on the normalization constant.

\subsection{Data-driven Langevin equation}\label{sec:ddLE}
\begin{figure}
  \centering
  \includegraphics[width=.9\linewidth]{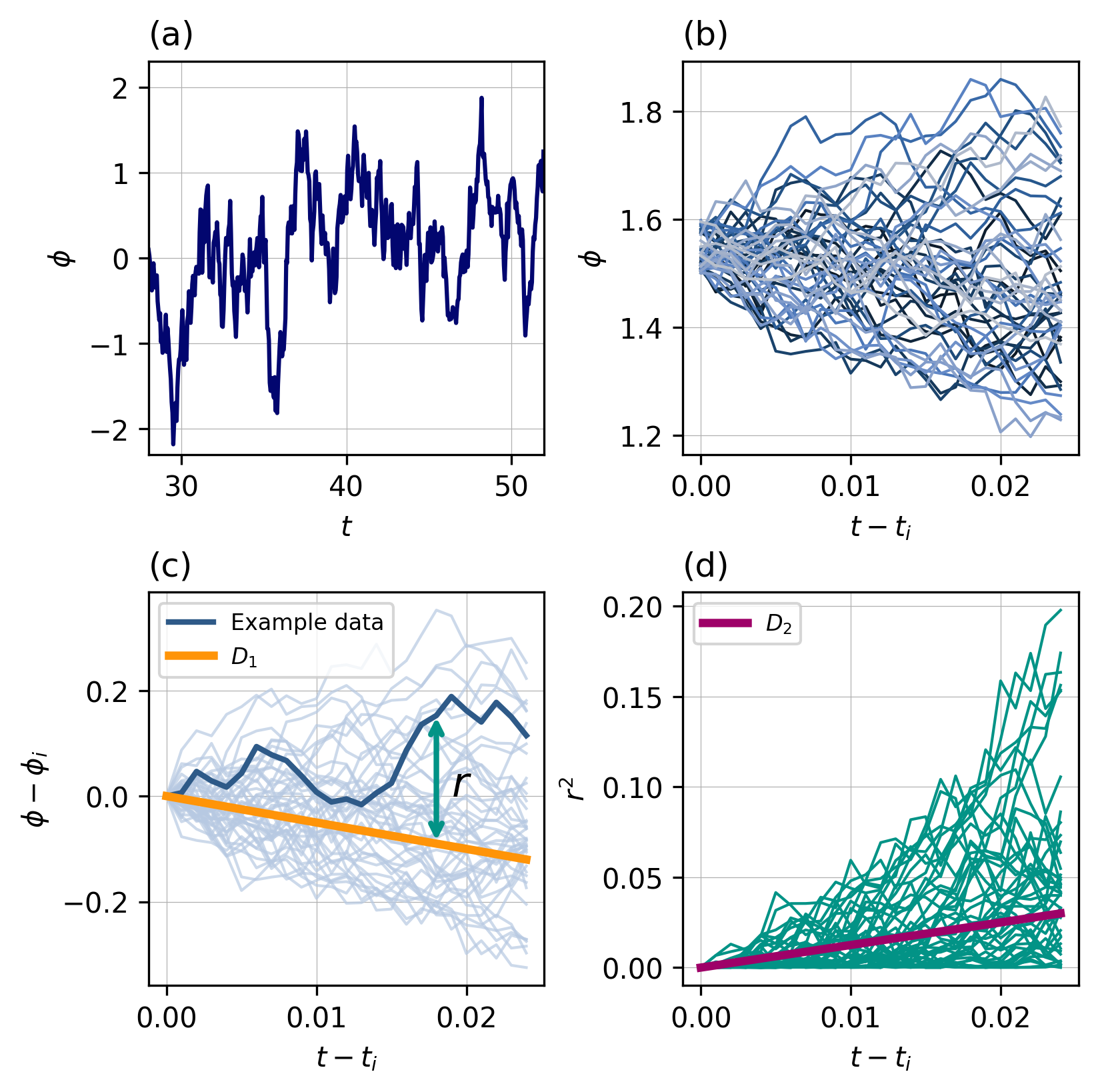}
  \caption {An illustration of the procedure of data-driven LE to diagnose the drift ($D_1$) and diffusion ($D_2$) coefficients. (a) The analysis is performed for sections of the time series data generated from solving LE (Eq.~\ref{eq:LE}) with D$_1 = -\phi$ and $D_2 = 1$. The units of time are arbitrary. (b) To obtain the coefficients, time series segments with a certain duration (here $\Delta t = 0.025$) are selected starting from $t_i$ where $\phi(t_i)$ is within a specified range (here $\phi(t_i) \in \langle 1.5, 1.6]$). (c) $D_1$ corresponds to the mean gradient of the selected data (slope of the bold orange line). A residual ($r(t-t_i)$, teal double-arrow) can be defined as the difference between the data (randomly selected dark blue line) and the mean fitted tendency ($D_1$, orange line). (d) $D_2$ (magenta line) can be diagnosed as half the mean slope of the $r^2$ series (teal lines).}
  \label{fig:f01}
\end{figure}

The state-dependent drift and diffusion coefficients can be estimated from time series data using the first and second order moments of the data increments. For illustrative purposes, a graphical depiction of the procedure of estimating the coefficients for synthetically generated data is given in Fig.~\ref{fig:f01}. The reconstruction of drift and diffusion is based on the second order truncation of the Kramers--Moyal expansion \citep[cf.~Appendix \ref{sec:app};][]{rahimitabarAnalysisDataBasedReconstruction2019,araniExitTimeMeasure2021}. Higher-order terms of the expansion need to be negligible for LE to be valid. Mathematical checks on the higher-order moments and coefficients, such as Pawula and Wick's theorem, help validate the suitability of this model (see Appendix \ref{sec:app}).    

\section{Fluctuation-induced bimodality in Arctic boundary-layer clouds}
\subsection{The longwave radiation ratio features a stable state at the clear-cloudy transition}\label{sec:netRad}
\begin{figure}[hbt!]
\centering
  \includegraphics[width=.9\linewidth]{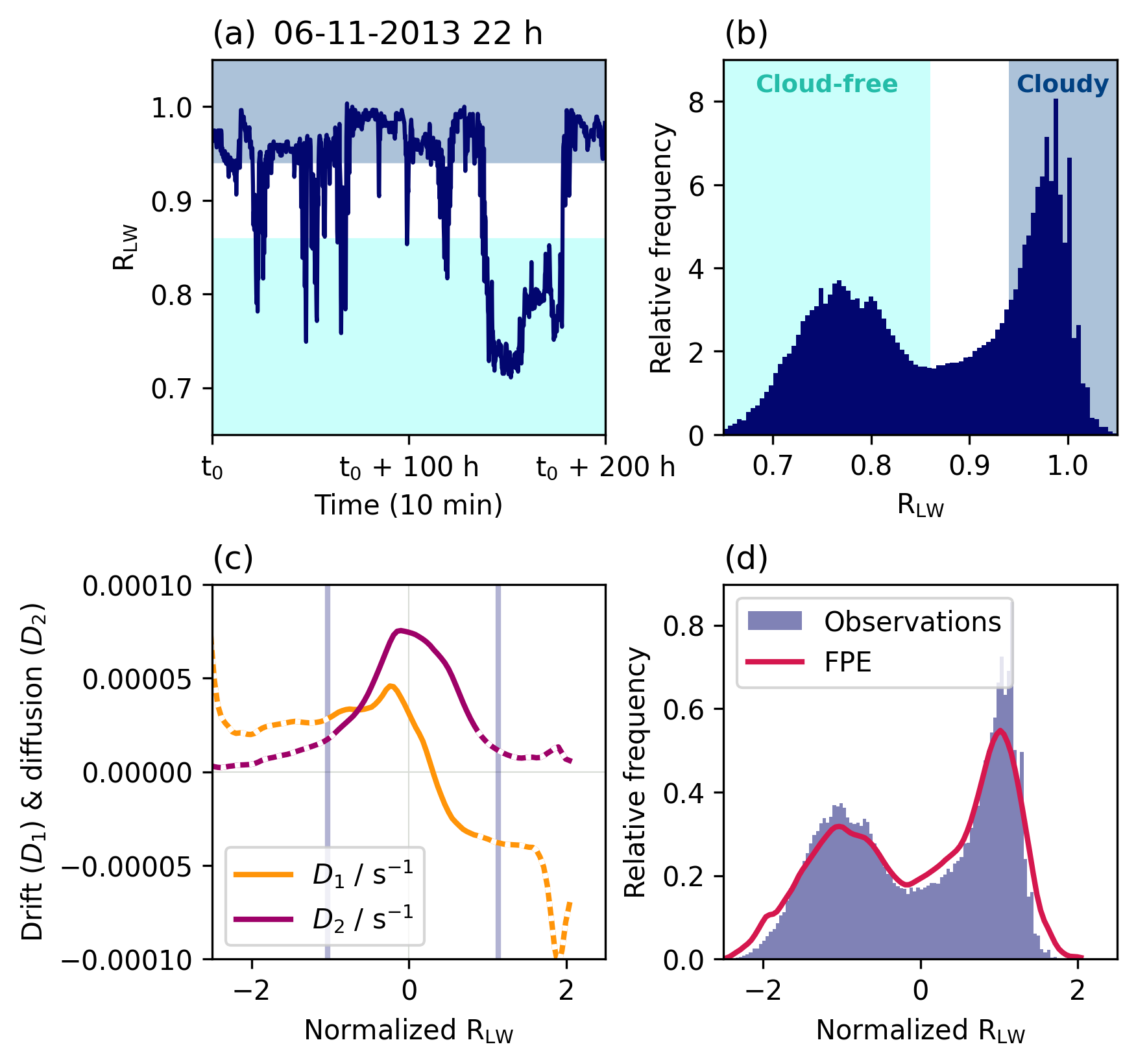}
  \caption {Time series characteristics, empirical probability density, drift-diffusion coefficients of longwave radiation ratio (R$_{\text{LW}}$), and the steady-state solution of FPE. (a) A 200-h representative segment of the R$_{\text{LW}}$ time series starting from 6 November 2013 22 UTC with shaded regions marking "Cloud-free" (R$_{\text{LW}} \leq 0.85$, light cyan) and "Cloudy" (R$_{\text{LW}} \geq 0.95$, dark blue) regimes. (b) Bimodal relative frequency distribution of R$_{\text{LW}}$ across the full dataset. (c) Drift ($D_1$, orange) and diffusion ($D_2$, magenta) coefficients as a function of normalized R$_{\text{LW}}$. The dashed segments indicate where our system deviates from Markovianity (see Appendix \ref{sec:app1}). Purple vertical lines indicate the two local maxima of the histogram in (d). (d) Steady-state solution of FPE (Eq.~\ref{eq:fpe_s}) using the diagnosed $D_1$ and $D_2$ coefficients. The histogram of the normalized observations is in the background.}
  \label{fig:intro}
\end{figure}

An example of the observed time series (Fig.~\ref{fig:intro}a) supports the previously reported finding of an `on/off' behavior, where cloudy states (R$_\mathrm{{LW}} \geq 0.95$) persist for a few hours up to several days. This `on/off' behavior of Arctic clouds is also reflected in the histogram of R$_\mathrm{{LW}}$, which is bimodal (Fig.~\ref{fig:intro}b). The data classified as cloud-free (R$_\mathrm{{LW}} \leq 0.85$) are distributed over a relatively large range in R$_\mathrm{{LW}}$-space, compared to the data distribution in the cloudy regime (R$_\mathrm{{LW}} \geq 0.95$). 

We diagnose the state-dependent drift and diffusion coefficients, $D_1$(R$_\mathrm{{LW}}$) and $D_2$(R$_\mathrm{{LW}}$), respectively, from the time series R$_\mathrm{{LW}}(t)$ (Sect.~\ref{sec:dd_frame}). The dynamical steady states can be inferred from the zero-crossing of the drift parameter (Fig.~\ref{fig:intro}c). For a bistable system, there exist \textit{two} locations where the drift changes sign from positive to negative values with increasing R$_\mathrm{{LW}}$. Our analysis only identifies \textit{one} stable state near the mean of the time series data (Fig.~\ref{fig:intro}c). Even without considering a reconstruction of coefficients outside the transition region, where results might be affected by lack of clear Markovianity (dashed lines in Fig.~\ref{fig:intro}, see Appendix~\ref{sec:app}), it is reasonable to deduce that the central stable state is the only stable state. If there were additional stable states, e.g., at the location of the maxima of the bimodal distribution, this would by definition require the existence of two further unstable states in between each pair of stable states. We have, however, no indication of such a complex structure.
The central steady state thus indicates that the observed bimodality in the distribution (Fig.~\ref{fig:intro}b) is not driven by deterministic drift, but by state-dependent diffusion (see Fig.~\ref{fig:sketch}). The local maximum in $D_2(\mathrm{R}_{\mathrm{LW}})$ signifies a region of enhanced stochastic fluctuations, where large random steps rapidly drive the system away from its stable point. The reduced residence time in intermediate states creates a diffusion-induced valley, effectively splitting the probability density into two regimes. 
The stationary PDF reconstructed from the FPE specified by the reconstructed coefficients (Eq.~\ref{eq:fpe_s}) shows the expected bimodality (Fig.~\ref{fig:intro}d). This reconstruction remains robust for an out-of-sample test with the first 10 years as a training dataset and the remaining 5 years as a validation dataset (not shown), giving further confidence in our reconstructed coefficients.

%\subsection{Non-Markovian effects in observations}\label{sec:stoch_model}
\subsection{Multiscale nature of Arctic boundary-layer clouds}\label{sec:stoch_model}
\begin{figure}[hbt!]
  \centering
  \includegraphics[width=.7\linewidth]{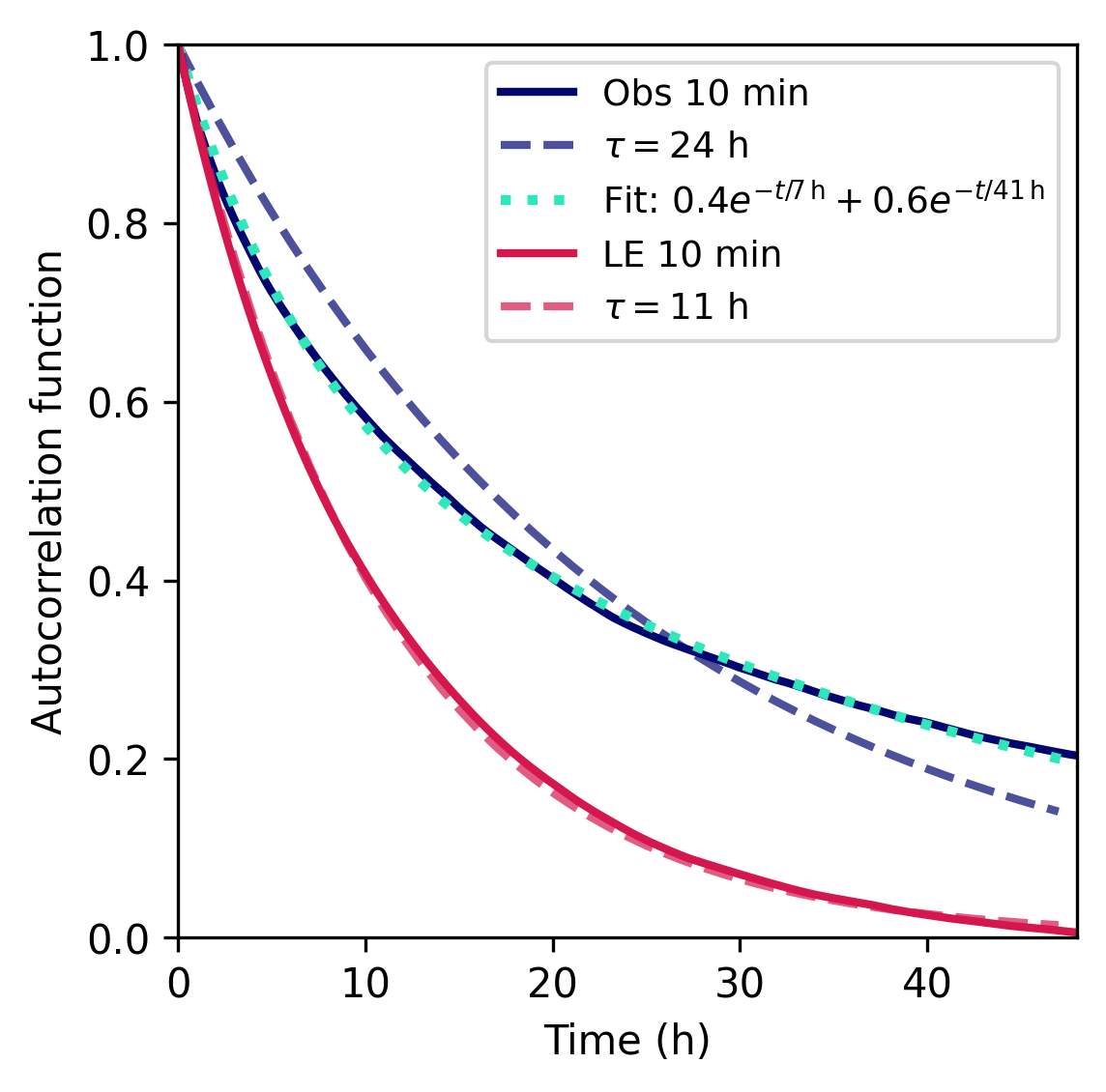}
  \caption {The autocorrelation function (ACF) for the time series data R$_\mathrm{{LW}}(t)$ from observations (dark blue) and reconstructed data using LE (Eq.~\ref{eq:LE}, red) for lag times up to 48\,h. The modeled data was generated from the diagnosed drift and diffusion coefficients for 1000 trajectories, integrated over 200\,h with a timestep of 1\,s. Additionally, two exponentially-decaying functions of the form $y = e^{-\frac{t}{\tau}}$ are shown (dashed), where $\tau$ is a timescale and its values are indicated in the legend ($\tau \in \{11, 24\}\,$h). These values were hand-tuned to approximately fit the ACFs. The observed ACF is fitted with a double exponential equation (dotted line).}
  \label{fig:acf}
\end{figure}

To evaluate the suitability of LE to describe the dynamics of the radiation ratio R$_\mathrm{{LW}}(t)$, we compare the autocorrelation functions (ACFs) of the observed and reconstructed time series (Fig.~\ref{fig:acf}). The latter consists of 1000 modeled trajectories obtained by solving LE (Eq.~\ref{eq:LE}) using the diagnosed drift and diffusion coefficients, a timestep of 1\,s with the Euler-Maruyama integration scheme, and reflecting boundary conditions. Increasing the number of trajectories to 10,000 and reducing the timestep to 0.01\,s did not change the reconstructed ACF. 

The observed ACF suggests the presence of multiple timescales, as it first decays faster ($\tau \approx 7~\mathrm{h}$) than the overall e-folding timescale of $\tau \approx 24~\mathrm{h}$ and then transitions to a slower decay ($\tau \approx 41~\mathrm{h}$) as indicated by the fitted double exponential decay. The shorter timescale could reflect the thermodynamic adjustment of boundary-layer variables, that, e.g. for subtropical stratocumulus, was found to take around 9\,h \citep{schubertMarineStratocumulusConvection1979a}. A possible explanation for the longer timescale is the passage of moist air intrusions \citep{woodsRoleMoistIntrusions2016} or polar low pressure systems \citep{moreno-ibanezRecentAdvancesPolar2021}, which persist for one to three days. In contrast, the reconstructed ACF decays exponentially with a characteristic timescale of  $\tau \approx 11~\mathrm{h}$. The Markovian Langevin model is calibrated to reproduce the local one-step statistics of the increments but does not account for longer temporal correlations as they might arise from variability in the synoptic forcing. Consequently, it underestimates the persistence of the observed time series and decorrelates faster. In addition to the effects of slow synoptic variability, which can be considered a source of non-stationarity, the reconstructed ACF might also reflect some effects of non-Markovianity, as we identify for the segments outside of the transition region (see Fig.~\ref{fig:intro}c and Appendix \ref{sec:app}).

\subsection{Location of stable point varies with synoptic forcing}\label{sec:syn}
Given the possible imprint of the synoptic forcing discussed in the previous section, we investigate how the synoptic conditions influence the statistics of the flux ratio (R$_\mathrm{{LW}}$). For this purpose, we repeat the analysis by applying LE (Eq.~\ref{eq:LE}) to R$_\mathrm{{LW}}(t)$, which now is binned for synoptic conditions retrieved from ERA5.

\begin{figure*}
  \centering
  \includegraphics[width=.9\textwidth]{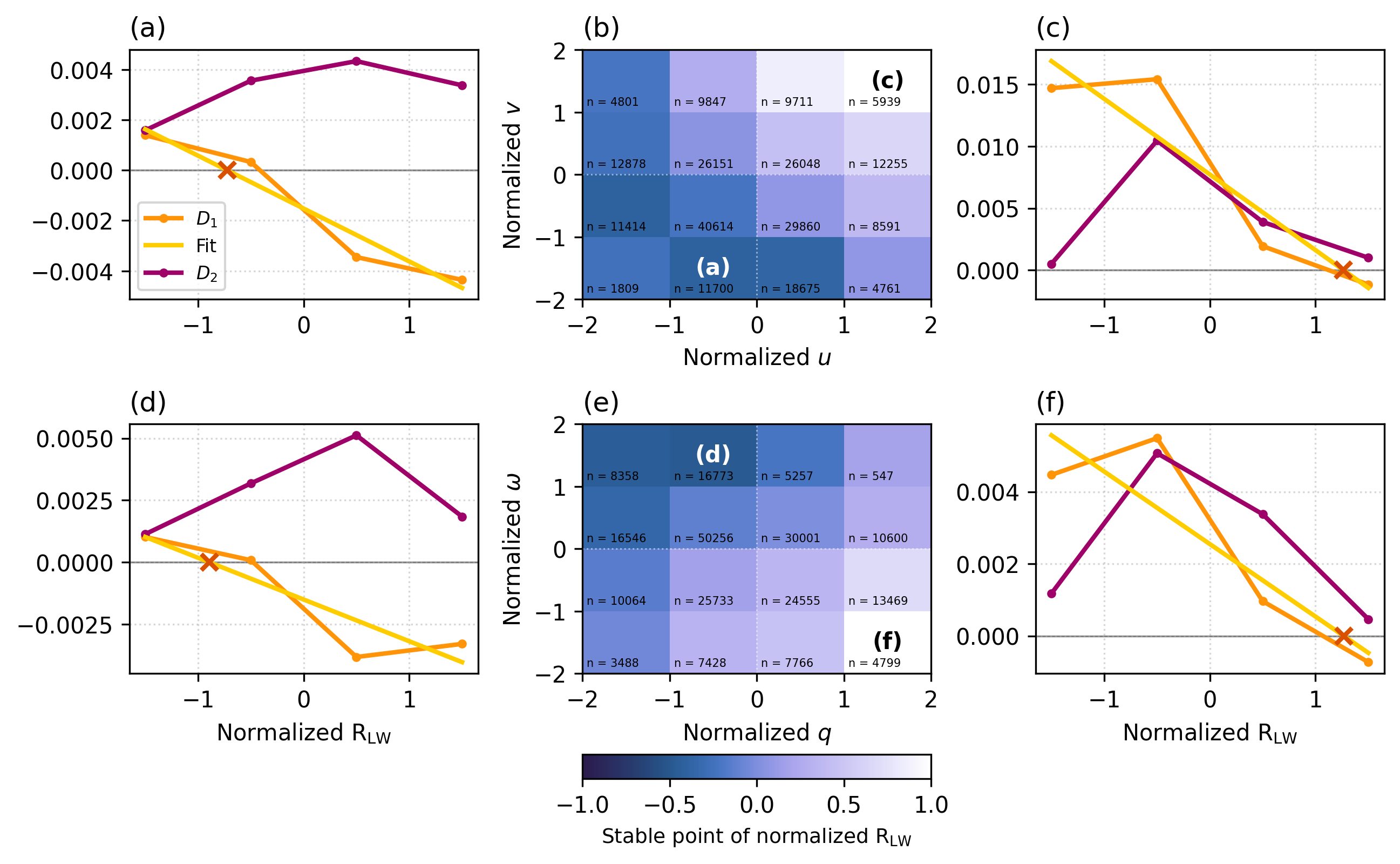}
  \caption {Langevin analysis showing stable fixed points and drift-diffusion profiles stratified by meteorological conditions. (a, c) Profiles of drift $D_1$ (orange), linear fit to $D_1$ (yellow), and diffusion $D_2$ (magenta) as a function of normalized R$_{\text{LW}}$, with the zero-crossing marked by a red `$\times$', evaluated at bins marked as (a) and (c) in (b). (b) Two-dimensional grid map of stable fixed points ($D_1 = 0$) of normalized R$_{\text{LW}}$ projected across normalized $u$ and $v$ wind components. Numbers in each bin indicate the number of data points. (d, f) Corresponding $D_1$ and $D_2$ profiles evaluated at locations marked as (d) and (f) in panel (e). (e) Two-dimensional grid map of stable fixed points of normalized R$_{\text{LW}}$ projected across normalized $q$ (specific humidity) and $\omega$ (vertical velocity). The shared colorbar indicates the position of the stable fixed point along the normalized R$_{\text{LW}}$ axis.}
  \label{fig:le_era}
\end{figure*}

First, we select the horizontal wind components $u$ and $v$ at the 900\,hPa pressure level as a proxy for the local synoptic conditions (Fig.~\ref{fig:le_era}a--c). We see that the location of the steady state shifts from negative values (i.e., R$_{\mathrm{LW,norm}} \approx -0.9$ for cloud-free conditions in Fig.~\ref{fig:le_era}a) to positive values (i.e., R$_{\mathrm{LW,norm}} \approx 1.1$ in Fig.~\ref{fig:le_era}c) for cloudy conditions. Moreover, cloud-free conditions appear with northeasterly winds, while cloudy conditions are associated with southwesterly winds. \citet{gierensLowlevelMixedphaseClouds2020} also reported a higher occurrence of low-level clouds with westerly winds for Ny-Ålesund. Southwesterly winds are associated with moist intrusions into the Arctic region, thus promoting cloud formation \citep{maturilli2013climatology,woodsRoleMoistIntrusions2016}. 

We conduct the same analysis for a $q, \omega$ parameter space (Fig.~\ref{fig:le_era}d$-$f). Here $q$ represents the specific humidity in the atmosphere and $\omega$ is the vertical pressure velocity. Negative values of $\omega$ correspond to upward motions and vice-versa. As expected, we find cloudy conditions for high $q$ and low $\omega$, which are indicators for cloud formation. For low $q$ and high $\omega$ cloud-free conditions are favored, as drier air and subsidence inhibit cloud formation and lead to the dissipation of existing clouds. 

The results from the horizontal wind analysis are not independent from those of the $\omega$ and $q$ parameters. To investigate the co-variability, we perform a principal component analysis on the $u,v,\omega$ and $q$ data. The first principal component
\begin{equation}\label{eq:PCA}
  P_1 = 0.20u + 0.61v - 0.50\omega + 0.59q.
  %P_2 = -0.81u + 0.32v - 0.40w + 0.29q.
\end{equation}
accounts for 47\,\% of the total variance. It confirms that the meteorological conditions are correlated with each other. Cloud-favorable conditions are found with poleward (positive $u$ and $v$) moisture transport (positive $q$) alongside uplifting motion (negative $\omega$), which generates the required supersaturation for cloud formation. Studies also highlighted the importance of surface pressure on the two radiative states, where relatively weaker high-pressure systems favor the cloudy state, while stronger high-pressure systems are associated with the cloud-free state \citep{zuidemaArcticSpringtimeMixedPhase2005,morrisonResiliencePersistentArctic2012,lacWeakInfluenceSurface2026}.

%\subsection{Illustrative example for the emergence of a bimodal distribution from state-dependent noise}
\subsection{Diffusion as a result of sub-synoptic-scale processes}\label{sec:toymodel}
The changing positions of the steady state due to synoptic variability (Fig.~\ref{fig:le_era}) modify the global diffusion coefficient (Fig.~\ref{fig:intro}c). Individual trajectories relax towards their respective stable state, which deviates from the global steady state. When we do not stratify by synoptic conditions, the resulting additional variability enhances the global diffusion coefficient. The diffusion remains state-dependent when controlling for the synoptic conditions (Fig.~\ref{fig:le_era}). This indicates that processes on smaller than synoptic scales, i.e., especially mesoscale boundary-layer and microscale processes, lead to the increased variability in the transition region.

\begin{figure}
  \centering
  \includegraphics[width=.9\linewidth]{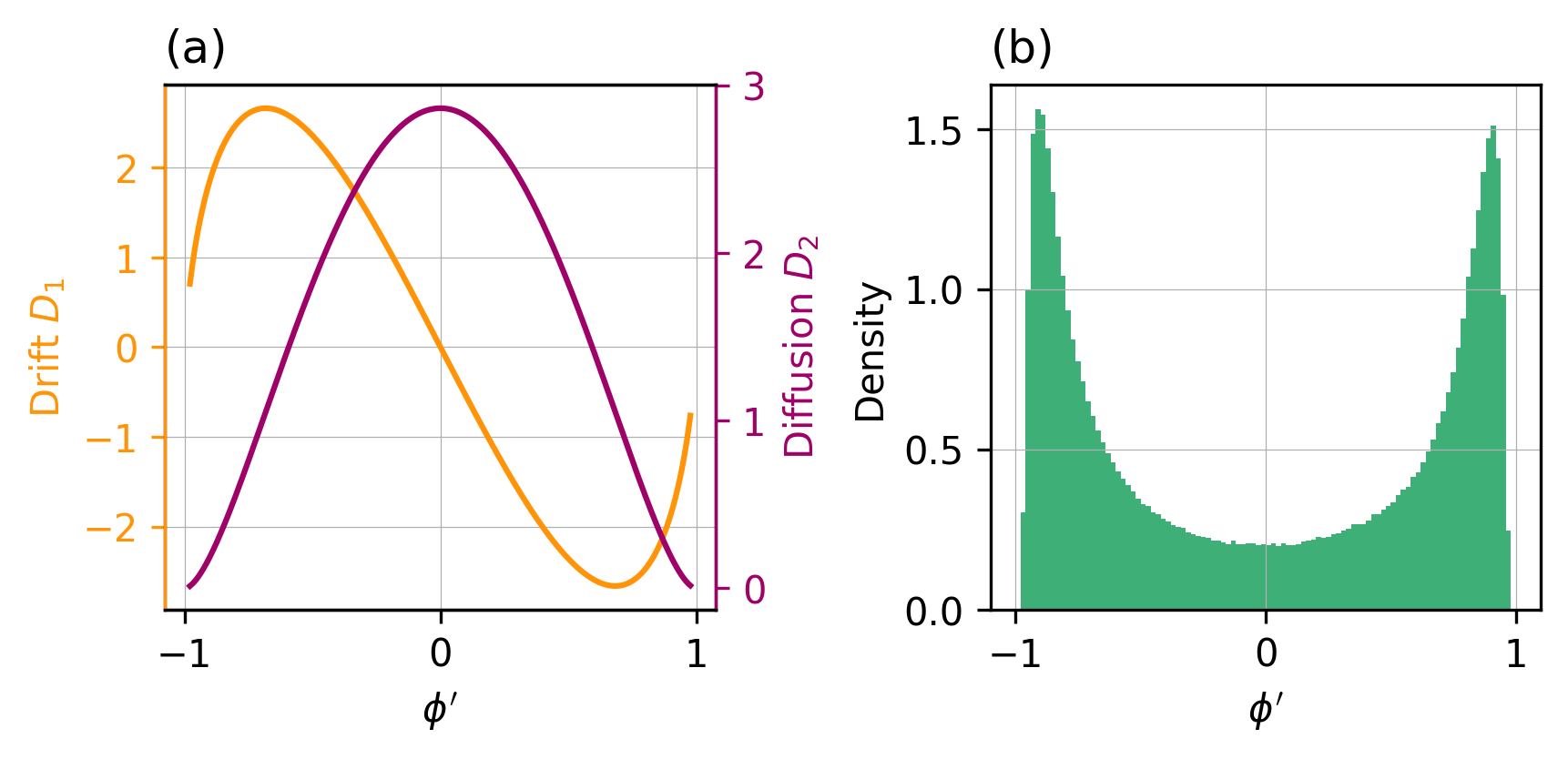}
  \caption {Illustrative example. (a) Drift ($D_1$, orange) and diffusion ($D_2$, magenta) coefficients for $\phi' = \mathrm{erf}\left(a \phi \right)$ with $a=1.5$ and the resulting PDF (b).}
  \label{fig:toymod}
\end{figure}

We further support this argument by illustrating how state-dependent variability in the radiative state of the Arctic boundary layer can result from combining a microscale non-linearity like cloud formation with uniform variability of a boundary-layer bulk quantity like humidity.
%This section illustrates how the observed bimodality in Arctic clouds and the accompanying diffusion profile, that peaks at the position of the minimum in the histogram, arise. We exemplify this by showing how a bimodal distribution can originate from a ``hidden'' underlying unimodal process.
%
Consider a Langevin process $\phi(t)$, with $D_1(\phi) = -\phi$ and $D_2 = 1$. It can be verified that the stationary solution to FPE is a Gaussian PDF: $\rho_s \propto \exp(-\phi^2/2)$. We introduce a transformation $\phi' = \mathcal{L}\left(\phi\right)$ that maps a location in $\phi$ space to $\phi'$ space. This transformation marks the difference between a measured physical quantity $\phi'$, e.g., R$_\mathrm{{LW}}$ as proxy for the cloud state of the Arctic boundary layer, and the ``hidden'' quantity that is more closely related to the underlying processes $\phi$, e.g., boundary-layer humidity. We choose the Gaussian error function as a non-linear relationship between both spaces:
\begin{equation}\label{eq:trans}
  \phi' = \mathrm{erf}\left(a \phi \right).
\end{equation}
For our example, this could capture the sudden onset of cloud formation once boundary-layer humidity reaches saturation.
The drift and diffusion coefficients that describe the $\phi'(t)$ time series differ from those associated with $\phi(t)$. In fact, we can compute them from the differential of the transformation \citep[Itô's lemma, e.g., ][]{riskenFokkerPlanckEquationMethods1996} 
\begin{equation}
  \mathrm{d}\phi' = \frac{\mathrm{d}\phi'}{\mathrm{d}\phi}\mathrm{d}\phi +\frac{1}{2} \frac{\mathrm{d}^2\phi'}{\mathrm{d}\phi^2}(\mathrm{d}\phi)^2.
\end{equation}
The LE for $\phi'$ becomes 
\begin{equation}
\begin{aligned}
  \mathrm{d}\phi' &= \left( D_1\left(\mathcal{L}^{-1}(\phi'\right) \frac{\mathrm{d}\phi'}{\mathrm{d}\phi}+D_2\left(\mathcal{L}^{-1}(\phi')\right)\frac{\mathrm{d}^2\phi'}{\mathrm{d}\phi^2}\right)\mathrm{d}t \\&+ \sqrt{2\left(\frac{\mathrm{d}\phi'}{\mathrm{d}\phi}\right)^2 D_2\left(\mathcal{L}^{-1}(\phi')\right)} \mathrm{d}W %\equiv D^{\phi'}_1\left(\phi'\right) \mathrm{d}t + \sqrt{2 D^{\phi'}_2\left(\phi'\right)} \mathrm{d}W.
\end{aligned}
\end{equation}
and we identify the new drift $D_1^{\phi'}$ and diffusion $D_2^{\phi'}$ coefficients:
\begin{align}
D^{\phi'}_1\left(\phi'\right) &= -\frac{2\left(2 a^2 + 1\right)}{\sqrt{\pi}}\operatorname{erf}^{-1}{(\phi')}e^{- \operatorname{erf}^{-2}{(\phi')}},\\  
  D_2^{\phi'}\left(\phi'\right) &= \frac{4a^2}{\pi} e^{-2\mathrm{erf}^{-2}(\phi')}.
\end{align}
In Fig.~\ref{fig:toymod}, we show the drift and the diffusion coefficients for $\phi'(t)$ with $a = 1.5$.  We see that the diffusion parameter $D_2^{\phi'}(\phi')$ has a maximum at the location of the steady state ($\phi' = 0$). The resulting histogram shows a bimodal structure, i.e., the transformation induces a bimodality from the underlying unimodal Gaussian process.

\section{Discussion and conclusion}\label{sec:concl}  %% \conclusions[modified heading if necessary]
In this study, we apply a drift--diffusion framework to wintertime long-term observations of longwave radiation at Ny-Ålesund. As identified by prior studies across different Arctic measurement stations and measurement campaigns \citep{stramlerSynopticallyDrivenArctic2011,morrisonResiliencePersistentArctic2012,raddatzDownwellingLongwaveRadiation2015,bertossaTwoUbiquitousRadiative2024,dahlkeTwoArcticWintertime2025}, the net longwave radiation exhibits a bimodal distribution. We express the net longwave radiation through the atmospheric longwave flux ratio R$_\mathrm{{LW}}$ and confirm this bimodality using a 15-year observational record. Our analysis of R$_\mathrm{{LW}}$ shows that the bimodal distribution of Arctic cloudiness is not explained by a drift contribution as it would result from a double-well potential. Instead, the drift exhibits a single stable point. At the same time, the diffusion is strongest close to this stable point. In this region, the strong diffusion produces large stochastic fluctuations, which move the system towards either the cloud-free or cloudy regime. Our results show further that the location of the steady state in R$_\mathrm{{LW}}$ space can shift depending on synoptic conditions, while diffusion remains state-dependent. Favorable conditions for the cloudy state include southwesterly winds, which may bring moisture, and upward motion. Our findings align with independent work by \citet{limStochasticCloudFluctuations2026}, who  performed an almost identical approach focusing on surface measurements from Utqiaġvik, Alaska. They validated their findings by also including surface measurements from Ny-Ålesund and ship measurements during the MOSAiC campaign \citep{shupeOverviewMOSAiCExpedition2022}. 

We formally illustrate how the observed bimodality in Arctic clouds and the accompanying diffusion profile that peaks at the position of the minimum in the histogram can be the result of a nonlinear process. In this interpretation, cloud formation converts fluctuations in boundary layer humidity into state-dependent fluctuations in the observed radiation, thereby linking microscale cloud-forming processes to variability in radiative properties at larger scales. \citet{limStochasticCloudFluctuations2026} provide a possible bridge between these scales by linking the inferred diffusion structure quantitatively to liquid water path variability and nonlinear radiative sensitivity.

The interpretation of these results needs to take several limitations into account. Methodologically, our 1D Langevin model assumes memoryless (Markovian) dynamics. In this framework, the true non-Markovian dynamics are projected onto a Markovian process that reproduces the one-step statistics of the increments (i.e., the conditional moments) at the chosen temporal resolution (i.e., $\tau = 10\,$min), which roughly corresponds to a boundary-layer eddy-turnover time. Effects of slow-synoptic-scale processes and memory, which is especially present outside of the transition region (as identified by the Chapman-Kolmogorov test, Appendix~\ref{sec:app}), are not explicitly captured. Instead, they are absorbed into effective drift and diffusion coefficients. While this approximation captures the local dynamics (Fig.~\ref{fig:intro}), it does not preserve long-time temporal correlations as illustrated by the ACF (Fig.~\ref{fig:acf}). \citet{limStochasticCloudFluctuations2026} also mention the possible presence of memory because their Chapman-Kolmogorov tests deteriorate for decreasing temporal resolution.

Our observational dataset feature geographical and local environmental constraints. As our dataset is confined to a single location (Ny-Ålesund), spatial variability across broader Arctic regions and diverse surface types, such as open ocean versus sea ice, is not captured. In addition, the local topography around Ny-Ålesund heavily influences atmospheric conditions by triggering localized cloud formation \citep{schemannSimulationMixedphaseClouds2020}, katabatic winds, and wind channeling effects  \citep{esauWindClimateKongsfjorden2012}. Such effects might contribute to memory effects. Reanalysis products furthermore struggle with fine-scale boundary layer structures, atmospheric inversions, and cloud microphysics in the Arctic, potentially smoothing out sharp state transitions \citep{pernovComparisonSelectedSurface2024}.

In summary, our analysis cautions against interpreting observed bimodality as an indication of an underlying deterministic dynamics that is bistable. In other words, the multiscale nature of the atmosphere may lead to interscale coupling. As already suggested by others \citep{morrisonResiliencePersistentArctic2012}, our analysis specifically suggests that microscale processes as well as processes on the scale of boundary-layer cloud fields are crucial to fully explain the statistics of the wintertime Arctic boundary layer.

%% The following commands are for the statements about the availability of data sets and/or software code corresponding to the manuscript.
%% It is strongly recommended to make use of these sections in case data sets and/or software code have been part of your research the article is based on.

%\codeavailability{TEXT} %% use this section when having only software code available

%\dataavailability{TEXT} %% use this section when having only data sets available

%\sampleavailability{TEXT} %% use this section when having geoscientific samples available

%\videosupplement{TEXT} %% use this section when having video supplements available

\newpage
\renewcommand\thesection{\Alph{section}}
\setcounter{section}{0}
\section{Appendix: Reconstruction procedure for Langevin dynamics from univariate time series}\label{sec:app}
We follow the reconstruction procedure provided by \citet{rahimitabarAnalysisDataBasedReconstruction2019} to ensure that the Langevin equation can be applied to our data. The following steps outline the procedure accompanied with analysis figures. 

\subsection{Verification of Markovianity of time series}
\label{sec:app1}
\begin{figure*}[hbt!]
  \centering
  \includegraphics[width=.9\linewidth]{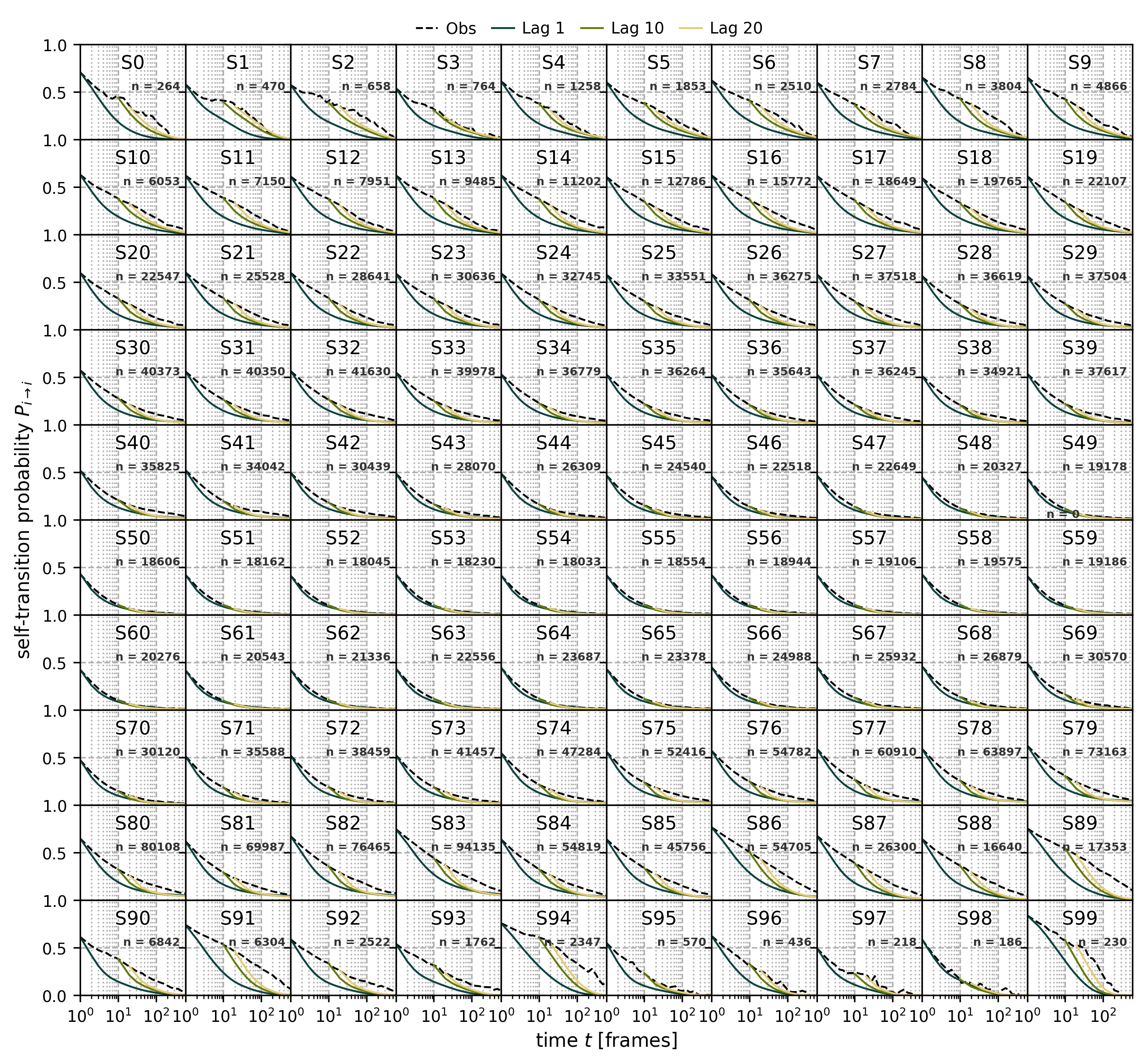}
  \caption {Chapman–Kolmogorov validation test for a 100-state discretized Markov model. Each subplot displays the self-transition probability $P_{i \to i}(t)$ as a function of time $t$ (in frames) for individual microstates (S0–S99). Black dashed lines represent the directly observed estimates (Obs), while solid colored curves show the predictions calculated at lag times $\tau \in \{1, 10, 20\}$ frames. Sample counts $n$ within each subplot indicate the total number of transitions originating from state $S_i$.}
  \label{fig:chapmankolmo}
\end{figure*}

\begin{figure}
  \centering
  \includegraphics[width=.9\linewidth]{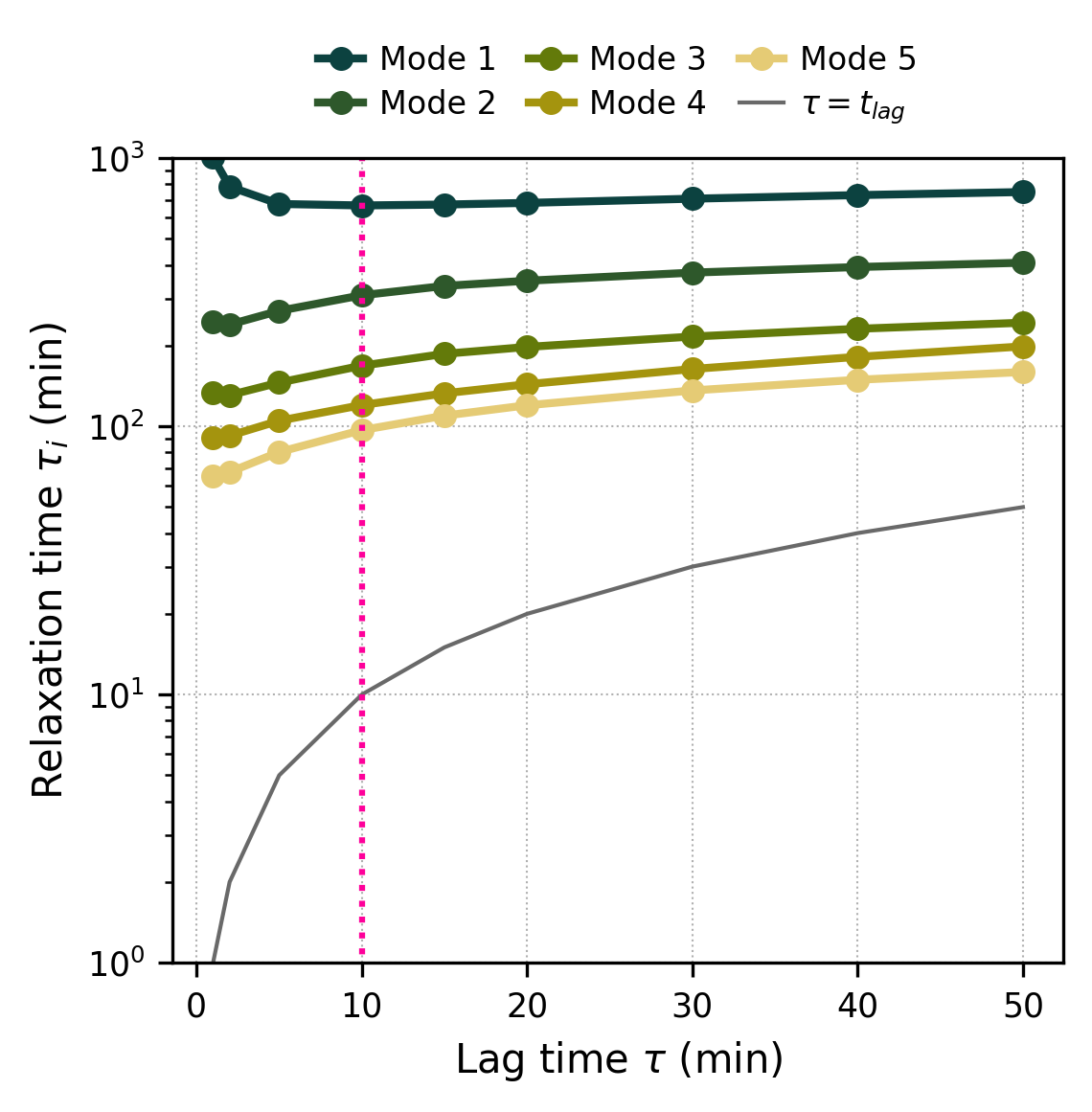}
  \caption {Relaxation times as a function of lag time $\tau$. The five slowest relaxation timescales ($\tau_i$, Modes 1–5) estimated from the 100-state Markov model are plotted against lag times ranging from $\tau = 1$ to $50$ min. The gray solid line indicates the theoretical resolution limit ($\tau = t_{\text{lag}}$). The vertical magenta dotted line highlights the selected Markov–Einstein timescale ($\tau = 10$ min), where non-Markovian memory effects decay and the relaxation modes enter stable plateaus.}
  \label{fig:impliedts}
\end{figure}

The Langevin equation assumes that the system's future state depends only on the present state and not on its history. Hence, we need to check if our system has memory. We verify the Markovianity of our data on the basis of the Chapman Kolmogorov equation using the python package \texttt{msmhelper} \citep{nagelMsmhelperPythonPackage2023}. For any system, we can estimate a Markovian state model. This is a statistical model that describes transitions between discrete states as a Markov (memoryless) process. The dynamics are then given by a transition matrix $T(\tau)$, where $T_{ij} (\tau)= P(s_t = j\vert s_{t-\tau}=i)=P_{i\rightarrow j}$ is the transition probability from state $i$ to state $j$ in the time step $\tau$. Figure \ref{fig:chapmankolmo} illustrates the probability of the system to stay within a given state $S$ after time $t$ for different lag times (1, 10, and 20\,min). We choose 100 states corresponding to  the 100 bins we used to discretize our data. A system is Markovian if the predicted probabilities (colored lines) match the observed probabilities (black, dashed line). We see that for the states S36 to S74 we achieve an agreement between the prediction and observation, indicating that the coarse-grained dynamics in this region are approximately Markovian. For the remaining states, however, the prediction strongly deviates from the observations, indicating that the system retains memory. We hypothesize that the strong diffusion present in the intermediate states leads to a faster de-correlation of trajectories, reducing the effective memory time. For this reason, we mark in Fig.~\ref{fig:intro}c the computed coefficients for bins with memory effects. Even though the Markovian assumption of LE does not hold in the outer segments, it is sufficient for our analysis to reconstruct the dynamics in the intermediate segments with the highest diffusion and steepest drift if we assume that a very complex dynamical structure with several stable and associated instable points is implausible (see main text). Thus, we believe it is an appropriate approximation to apply LE to investigate the nature of Arctic boundary-layer dynamics. Nonetheless, to more closely represent our data as a Markovian system, we identify the minimum lag time $\tau$ at which the relaxation times of the Markov model $t_i = -\tau / \ln \lambda_i$ (where $\lambda_i$ are the eigenvalues of the transition matrix $T(\tau)$) converge to constant values, i.e., become independent of the chosen lag time. Figure \ref{fig:impliedts} shows $t_i$ for the first five dynamical modes (corresponding to the five largest non-trivial eigenvalues of $T(\tau)$). Mode 1 is the slowest with a relaxation time of 700\,min at a lag time of 10\,min, after which it stabilizes. Before that, there is a pronounced initial decrease indicative of non-Markovian processes. For the other modes, the changes for lag times $<$ 10\,min are less pronounced. Based on this analysis, we resample our data to 10\,min for the subsequent analysis to minimize the effects of memory in the system.  

\subsection{Estimation of Kramers-Moyal coefficients}\label{sec:app2}
We use the python package \texttt{KramersMoyal} \citep{gorjaoKramersmoyalKramersMoyalCoefficients2019} to estimate the lag-$\tau$ moments $M_m(x,\tau)$ up to the 6\textsuperscript{th} order ($m \in \{1, 2, 3, 4, 5, 6\}$), where the first and second moments are used to compute the drift and diffusion coefficients, and higher order moments are needed to validate the use of LE. The implementation of the Kramers-Moyal expansion employs a kernel-based estimation, i.e., the Nadaraya-Watson kernel estimator, which is more robust against data sparsity and computationally more efficient than the binning approach \citep{gorjaoKramersmoyalKramersMoyalCoefficients2019}. The Kramers-Moyal coefficients are defined in the continuous limit as $D_m(x) =\frac{1}{m!} \lim_{\tau\rightarrow 0}\frac{M_m(x, \tau)}{\tau}$. For our discretely sampled data, we estimate the drift and diffusion coefficients (Fig.~\ref{fig:intro}) as 
\begin{equation}
    D_1\left(x\right) = \frac{M_1(x,\tau)}{\tau},\\  
    D_2\left(x\right) = \frac{M_2(x,\tau)-M_1(x,\tau)^2}{2\tau},
\end{equation}
where the subtraction of $M_1^2$ in the equation for $D_2$ removes the finite deterministic drift during the sampling interval. This provides only a partial correction of finite-time effects; a full finite-time correction would require additional terms involving derivatives of the drift and diffusion coefficients \citep{gorjaoArbitraryOrderFiniteTimeCorr2021}.

\subsection{Verification of vanishing higher-order terms}\label{sec:app3}
\begin{figure*}
  \centering
  \includegraphics[width=.9\linewidth]{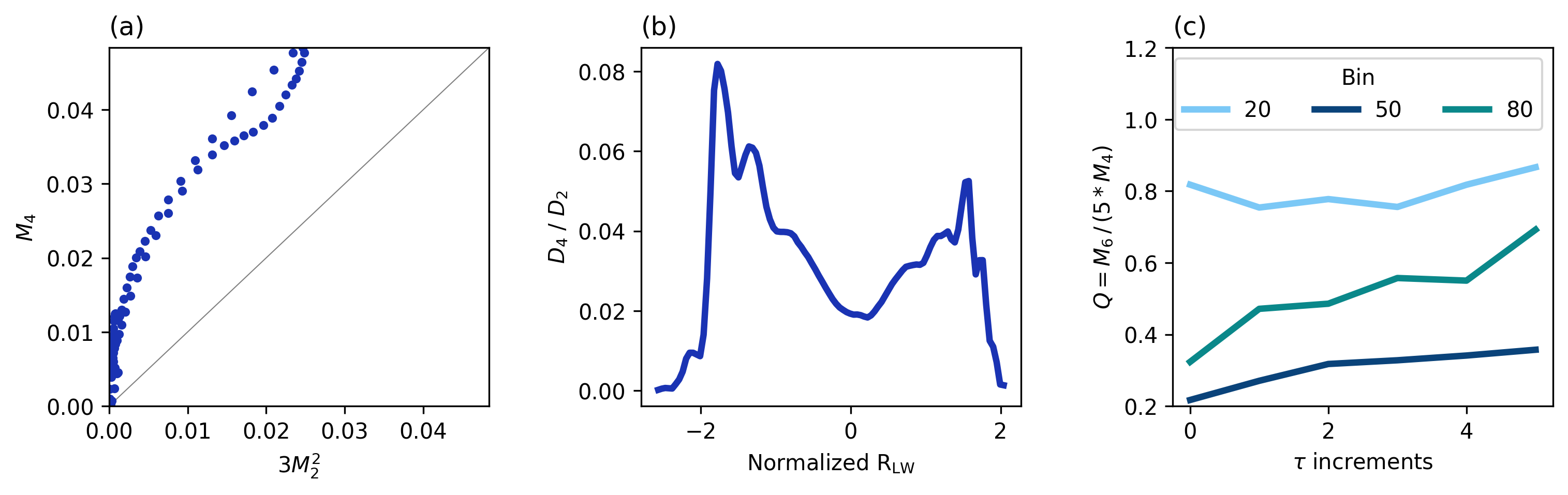}
  \caption {Verification of Gaussian increment statistics and Kramers–Moyal truncation at $\tau = 10$ min. (a) Scatter plot of the fourth conditional moment $M_4$ against the second conditional moment $3M_2^2$. The 1:1 gray reference line denotes ideal Gaussian behavior under Wick's theorem. (b) Ratio of the fourth- to second-order Kramers–Moyal coefficients ($D_4 / D_2$) across normalized longwave radiation (R$_{\text{LW}}$). (c) $Q = M_6 / (5 M_4)$ evaluated across representative state bins (bins 20, 50, and 80) over short $\tau = 10$\,min increments.}
  \label{fig:km_moms}
\end{figure*}

The Pawula theorem states that there are three possible outcomes for the Kramers-Moyal expansion: (i) the expansion is truncated at $m=1$, i.e., the system is deterministic, (ii) the expansion ends at $m=2$ resulting in the a drift--diffusion system described by the Langevin equation, or (iii) the expansion contains all terms up to $m=\infty$. To determine which outcome applies, we use Wick's theorem $M_4 = 3 M_2^2$, which holds for Gaussian increments and ensures that all orders $m\geq 3$ vanish. Figure \ref{fig:km_moms}a shows that our data points fall above the 1:1 line, which reflects the heavy-tailed behavior of our distribution and indicates that higher orders do not necessarily vanish. A second verification for vanishing higher orders is the ratio of $D_4$ and $D_2$ (Fig.~\ref{fig:km_moms}b). If the ratio is well below unity, it is justifiable for practical purposes to neglect higher orders \citep{araniExitTimeMeasure2021}. Finally, we check if jump processes, i.e., sudden changes in the time series, are present. For a diffusive process, the function Q$(\tau, x_0) = M_6(\tau, x_0)/(5 M_4(\tau,x_0))$, where $x_0$ is a certain bin, scales linearly with $\tau$. Figure \ref{fig:km_moms}c confirms a linear relationship; our system does not include jump processes and that higher orders can be neglected. Thus, it can be described by a drift and diffusion framework, i.e., the Langevin equation. Coarsening of the temporal resolution beyond 10\,min to further reduce memory effects resulted in stronger violations of Pawula and Wick's theorem.

% \noappendix       %% use this to mark the end of the appendix section. Otherwise the figures might be numbered incorrectly (e.g. 10 instead of 1).

%% Regarding figures and tables in appendices, the following two options are possible depending on your general handling of figures and tables in the manuscript environment:

%% Option 1: If you sorted all figures and tables into the sections of the text, please also sort the appendix figures and appendix tables into the respective appendix sections.
%% They will be correctly named automatically.

%% Option 2: If you put all figures after the reference list, please insert appendix tables and figures after the normal tables and figures.
%% To rename them correctly to A1, A2, etc., please add the following commands in front of them:

% \appendixfigures  %% needs to be added in front of appendix figures

% \appendixtables   %% needs to be added in front of appendix tables

%% Please add \clearpage between each table and/or figure. Further guidelines on figures and tables can be found below.

% \noindent \textit{Disclaimer.} During the final stages of preparing this manuscript, we became aware of related, independent, and as-yet-unpublished work by Jung-Sub Lim and Graham Feingold. %% optional section

\noindent \textit{Author contributions.} AvH preprocessed the data, conducted the initial analysis, and drafted an initial version of the manuscript. NO recomputed and refined the analysis, created the figures, and finalized the paper. VZ provided input on the mathematical concepts of the Langevin equation and contributed to the interpretation of the results. FG proposed, conceptualized, and supervised the project and provided input on the methodology and interpretation of the results. All authors contributed to the editing and reviewing of the manuscript. %% optional section

\noindent \textit{Acknowledgements.} AvH and FG acknowledge support from The Branco Weiss Fellowship - Society in Science, administered by ETH Zurich (2018-2024). FG and VZ acknowledge support by the European Union (ERC, MesoClou, 101117462). Views and opinions expressed are however those of the authors only and do not necessarily reflect those of the European Union or the European Research Council Executive Agency. Neither the European Union nor the granting authority can be held responsible for them. FG further acknowledges support by the Dutch Research Council through the Veni Talent Scheme (2018-2023). We thank Annica Ekman and Maura Dewey for the advice on the data sources used in this study. ChatGPT Pro was used for a pre-submission review, based on which the authors improved the clarity of the manuscript. No text or interpretations were machine-generated. Gemini was used for debugging parts of the analysis code. All code was reviewed and tested by the authors. The authors are fully aware that the use of AI does not change their author responsibilities.  %% optional section

\newpage
\footnotesize
\bibliography{main}

@article{araniExitTimeMeasure2021,
  title = {Exit Time as a Measure of Ecological Resilience},
  author = {Arani, Babak M. S. and Carpenter, Stephen R. and Lahti, Leo and {van Nes}, Egbert H. and Scheffer, Marten},
  year = 2021,
  journal = {Science},
  volume = {372},
  number = {6547},
  pages = {eaay4895},
  publisher = {American Association for the Advancement of Science},
  doi = {10.1126/science.aay4895}
}

@article{bakerBistabilityCCNConcentrations1990a,
  title = {Bistability of {{CCN}} Concentrations and Thermodynamics in the Cloud-Topped Boundary Layer},
  author = {Baker, Marcia B. and Charlson, Robert J.},
  year = 1990,
  journal = {Nature},
  volume = {345},
  number = {6271},
  pages = {142--145},
  publisher = {Nature Publishing Group},
  issn = {1476-4687},
  doi = {10.1038/345142a0},
  copyright = {1990 Springer Nature Limited}
}

@article{Bergeron1935,
      author = "T. Bergeron",
      title = "On the physics of clouds and precipitation",
      year = "1935",
      pages = "156–180",
      journal={Proc. 5th Assembly UGGI, Lisbon, Portugal, 1935}
}

@article{bernerStochasticParameterizationNew2017,
  title = {Stochastic {{Parameterization}}: {{Toward}} a {{New View}} of {{Weather}} and {{Climate Models}}},
  shorttitle = {Stochastic {{Parameterization}}},
  author = {Berner, Judith and Achatz, Ulrich and Batt{\'e}, Lauriane and Bengtsson, Lisa and de la C{\'a}mara, Alvaro and Christensen, Hannah M. and Colangeli, Matteo and Coleman, Danielle R. B. and Crommelin, Daan and Dolaptchiev, Stamen I. and Franzke, Christian L. E. and Friederichs, Petra and Imkeller, Peter and J{\"a}rvinen, Heikki and Juricke, Stephan and Kitsios, Vassili and Lott, Fran{\c c}ois and Lucarini, Valerio and Mahajan, Salil and Palmer, Timothy N. and Penland, C{\'e}cile and Sakradzija, Mirjana and von Storch, Jin-Song and Weisheimer, Antje and Weniger, Michael and Williams, Paul D. and Yano, Jun-Ichi},
  year = 2017,
  journal = {Bulletin of the American Meteorological Society},
  volume = {98},
  number = {3},
  pages = {565--588},
  publisher = {American Meteorological Society},
  issn = {0003-0007, 1520-0477},
  doi = {10.1175/BAMS-D-15-00268.1},
  chapter = {Bulletin of the American Meteorological Society}
}

@article{bertossaTwoUbiquitousRadiative2024,
  title = {Two {{Ubiquitous Radiative States Observed}} across the {{High Latitudes}}},
  author = {Bertossa, Cameron and L'Ecuyer, Tristan},
  year = 2024,
  journal = {Journal of Climate},
  volume = {37},
  number = {8},
  pages = {2585--2610},
  issn = {0894-8755, 1520-0442},
  doi = {10.1175/JCLI-D-23-0553.1},
  copyright = {http://www.ametsoc.org/PUBSReuseLicenses}
}

@article{chylekAnnualMeanArctic2022,
  title = {Annual {{Mean Arctic Amplification}} 1970--2020: {{Observed}} and {{Simulated}} by {{CMIP6 Climate Models}}},
  shorttitle = {Annual {{Mean Arctic Amplification}} 1970--2020},
  author = {Chylek, Petr and Folland, Chris and Klett, James D. and Wang, Muyin and Hengartner, Nick and Lesins, Glen and Dubey, Manvendra K.},
  year = 2022,
  journal = {Geophysical Research Letters},
  volume = {49},
  number = {13},
  pages = {e2022GL099371},
  issn = {1944-8007},
  doi = {10.1029/2022GL099371},
  copyright = {\copyright{} 2022. American Geophysical Union. All Rights Reserved.}
}

@article{dong10YearClimatology2010,
  title = {A 10 Year Climatology of {{Arctic}} Cloud Fraction and Radiative Forcing at {{Barrow}}, {{Alaska}}},
  author = {Dong, Xiquan and Xi, Baike and Crosby, Kathryn and Long, Charles N. and Stone, Robert S. and Shupe, Matthew D.},
  year = 2010,
  journal = {Journal of Geophysical Research: Atmospheres},
  volume = {115},
  number = {D17},
  issn = {2156-2202},
  doi = {10.1029/2009JD013489},
  copyright = {Copyright 2010 by the American Geophysical Union.}
}

@article{dahlkeTwoArcticWintertime2025,
  title = {The {{Two Arctic Wintertime Boundary Layer States}}: {{Disentangling}} the {{Role}} of {{Cloud}} and {{Wind Regimes}} in {{Reanalysis}} and {{Observations During MOSAiC}}},
  shorttitle = {The {{Two Arctic Wintertime Boundary Layer States}}},
  author = {Dahlke, Sandro and Rinke, Annette and Shupe, Matthew D. and Cox, Christopher J.},
  year = 2025,
  journal = {Atmospheric Science Letters},
  volume = {26},
  number = {4},
  pages = {e1298},
  issn = {1530-261X},
  doi = {10.1002/asl.1298},
  copyright = {\copyright{} 2025 The Author(s). Atmospheric Science Letters published by John Wiley \& Sons Ltd on behalf of Royal Meteorological Society.}
}

@article{driemel2018baseline,
  title={Baseline Surface Radiation Network (BSRN): structure and data description (1992--2017)},
  author={Driemel, Amelie and Augustine, John and Behrens, Klaus and Colle, Sergio and Cox, Christopher and Cuevas-Agull{\'o}, Emilio and Denn, Fred M and Duprat, Thierry and Fukuda, Masato and Grobe, Hannes and others},
  journal={Earth System Science Data},
  volume={10},
  number={3},
  pages={1491--1501},
  year={2018},
  publisher={Copernicus GmbH},
  doi={10.5194/essd-10-1491-2018}
}

@article{duffeyRepresentationArcticWinter2025,
  title = {Representation of {{Arctic Winter Atmospheric Boundary Layer Stability Over Sea Ice}} in {{CMIP6 Models}}},
  author = {Duffey, Alistair and Mallett, Robbie and Dutch, Victoria R. and Steckling, Julia and Hermant, Antoine and Day, Jonathan and Pithan, Felix},
  year = 2025,
  journal = {Journal of Geophysical Research: Atmospheres},
  volume = {130},
  number = {11},
  pages = {e2024JD041412},
  issn = {2169-8996},
  doi = {10.1029/2024JD041412},
}

@article{esauWindClimateKongsfjorden2012,
  title = {Wind {{Climate}} in {{Kongsfjorden}}, {{Svalbard}}, and {{Attribution}} of {{Leading Wind Driving Mechanisms}} through {{Turbulence-Resolving Simulations}}},
  author = {Esau, Igor and Repina, Irina},
  year = 2012,
  journal = {Advances in Meteorology},
  volume = {2012},
  number = {1},
  pages = {568454},
  issn = {1687-9317},
  doi = {10.1155/2012/568454},
  copyright = {Copyright \copyright{} 2012 Igor Esau and Irina Repina.}
}

@article{feingoldReversibilityTransitionsClosed2015,
  title = {On the Reversibility of Transitions between Closed and Open Cellular Convection},
  author = {Feingold, G. and Koren, I. and Yamaguchi, T. and Kazil, J.},
  year = 2015,
  journal = {Atmospheric Chemistry and Physics},
  volume = {15},
  number = {13},
  pages = {7351--7367},
  publisher = {Copernicus GmbH},
  issn = {1680-7316},
  doi = {10.5194/acp-15-7351-2015}
}

@article{Findeisen1938,
author = "Findeisen, Walter",
journal = "Meteorologische Zeitschrift",
year = "1938",
title = "{Kolloid-meteorologische Vorgänge bei der Niederschlagsbildung}",
volume = "55",
pages = "121-133"
}

@article{gierensLowlevelMixedphaseClouds2020,
  title = {Low-Level Mixed-Phase Clouds in a Complex {{Arctic}} Environment},
  author = {Gierens, Rosa and Kneifel, Stefan and Shupe, Matthew D. and Ebell, Kerstin and Maturilli, Marion and L{\"o}hnert, Ulrich},
  year = 2020,
  journal = {Atmospheric Chemistry and Physics},
  volume = {20},
  number = {6},
  pages = {3459--3481},
  publisher = {Copernicus GmbH},
  issn = {1680-7316},
  doi = {10.5194/acp-20-3459-2020}
}

@article{gorjaoKramersmoyalKramersMoyalCoefficients2019,
  title = {Kramersmoyal: {{Kramers--Moyal}} Coefficients for Stochastic Processes},
  shorttitle = {Kramersmoyal},
  author = {Gorj{\~a}o, Leonardo Rydin and Meirinhos, Francisco},
  year = 2019,
  journal = {Journal of Open Source Software},
  volume = {4},
  number = {44},
  pages = {1693},
  issn = {2475-9066},
  doi = {10.21105/joss.01693}
}

@Article{gorjaoArbitraryOrderFiniteTimeCorr2021,
AUTHOR = {Gorjão, Leonardo Rydin and Witthaut, Dirk and Lehnertz, Klaus and Lind, Pedro G.},
TITLE = {Arbitrary-Order Finite-Time Corrections for the Kramers–Moyal Operator},
JOURNAL = {Entropy},
VOLUME = {23},
YEAR = {2021},
NUMBER = {5},
ARTICLE-NUMBER = {517},
PubMedID = {33923154},
ISSN = {1099-4300},
DOI = {10.3390/e23050517}
}

@article{hersbach2020era5,
  title={The ERA5 global reanalysis},
  author={Hersbach, Hans and Bell, Bill and Berrisford, Paul and Hirahara, Shoji and Hor{\'a}nyi, Andr{\'a}s and Mu{\~n}oz-Sabater, Joaqu{\'\i}n and Nicolas, Julien and Peubey, Carole and Radu, Raluca and Schepers, Dinand and others},
  journal={Quarterly Journal of the Royal Meteorological Society},
  volume={146},
  number={730},
  pages={1999--2049},
  year={2020},
  publisher={Wiley Online Library},
  doi={10.1002/qj.3803}
}

@misc{hernandezAerosolMemoryStratocumulus2026,
  title = {Aerosol Memory in Stratocumulus Clouds Leads to Noise-Induced Patterns and Non-Ergodic Sampling},
  author = {Hernandez, Benjamin and Glassmeier, Franziska},
  year = 2026,
  number = {arXiv:2605.04002},
  eprint = {2605.04002},
  primaryclass = {physics.ao-ph},
  publisher = {arXiv},
  doi = {10.48550/arXiv.2605.04002},
  archiveprefix = {arXiv}
}

@article{heymsfieldContributionsLiquidIce2020,
  title = {Contributions of the {{Liquid}} and {{Ice Phases}} to {{Global Surface Precipitation}}: {{Observations}} and {{Global Climate Modeling}}},
  shorttitle = {Contributions of the {{Liquid}} and {{Ice Phases}} to {{Global Surface Precipitation}}},
  author = {Heymsfield, Andrew J. and Schmitt, Carl and Chen, Chih-Chieh-Jack and Bansemer, Aaron and Gettelman, Andrew and Field, Paul R. and Liu, Chuntao},
  year = 2020,
  journal = {Journal of the Atmospheric Sciences},
  volume = {77},
  number = {8},
  pages = {2629--2648},
  publisher = {American Meteorological Society},
  issn = {0022-4928, 1520-0469},
  doi = {10.1175/JAS-D-19-0352.1},
  chapter = {Journal of the Atmospheric Sciences}
}

@article{lacWeakInfluenceSurface2026,
  title = {Weak {{Influence}} of {{Surface Pressure}} on {{Arctic Radiative States From Winter}} to {{Spring Over}} the {{Sea-Ice}}},
  author = {Lac, Jean and Chepfer, H{\'e}l{\`e}ne and Shupe, Matthew D.},
  year = 2026,
  journal = {Geophysical Research Letters},
  volume = {53},
  number = {8},
  pages = {e2026GL122051},
  issn = {1944-8007},
  doi = {10.1029/2026GL122051},
  copyright = {\copyright{} 2026. The Author(s).}
}

@article{liResponseSimulatedArctic2017,
  title = {The {{Response}} of {{Simulated Arctic Mixed-Phase Stratocumulus}} to {{Sea Ice Cover Variability}} in the {{Absence}} of {{Large-Scale Advection}}},
  author = {Li, Zhujun and Xu, Kuan-Man and Cheng, Anning},
  year = 2017,
  journal = {Journal of Geophysical Research: Atmospheres},
  volume = {122},
  number = {22},
  pages = {12,335--12,352},
  issn = {2169-8996},
  doi = {10.1002/2017JD027086},
  copyright = {\copyright 2017. American Geophysical Union. All Rights Reserved.}
}

@article{limStochasticCloudFluctuations2026,
  title = {Stochastic Cloud Fluctuations Drive {{Arctic}} Winter Radiative Bistability},
  author = {Lim, Jung-Sub and Feingold, Graham},
  year = 2026,
  journal = {Science Advances},
  volume = {12},
  number = {40},
  pages = {eaeh3823},
  issn = {2375-2548},
  doi = {10.1126/sciadv.aeh3823}
}

@article{maturilli2013climatology,
  title={Climatology and time series of surface meteorology in Ny-{\AA}lesund, Svalbard},
  author={Maturilli, Marion and Herber, Andreas and K{\"o}nig-Langlo, Gert},
  journal={Earth System Science Data},
  volume={5},
  number={1},
  pages={155--163},
  year={2013},
  publisher={Copernicus Publications G{\"o}ttingen, Germany},
  doi={10.5194/essd-5-155-2013}
}

@article{mauritsenArcticCCNlimitedCloudaerosol2011,
  title = {An {{Arctic CCN-limited}} Cloud-Aerosol Regime},
  author = {Mauritsen, T. and Sedlar, J. and Tjernstr{\"o}m, M. and Leck, C. and Martin, M. and Shupe, M. and Sjogren, S. and Sierau, B. and Persson, P. O. G. and Brooks, I. M. and Swietlicki, E.},
  year = 2011,
  journal = {Atmospheric Chemistry and Physics},
  volume = {11},
  number = {1},
  pages = {165--173},
  publisher = {Copernicus GmbH},
  issn = {1680-7316},
  doi = {10.5194/acp-11-165-2011}
}

@article{moreno-ibanezRecentAdvancesPolar2021,
  title = {Recent Advances in Polar Low Research: Current Knowledge, Challenges and Future Perspectives},
  shorttitle = {Recent Advances in Polar Low Research},
  author = {{Moreno-Ib{\'a}{\~n}ez}, Marta and Laprise, Ren{\'e} and Gachon, Philippe},
  year = 2021,
  journal = {Tellus},
  volume = {73},
  number = {1},
  issn = {3035-9554},
  doi={10.1080/16000870.2021.1890412}
}

@article{morrisonResiliencePersistentArctic2012,
  title = {Resilience of Persistent {{Arctic}} Mixed-Phase Clouds},
  author = {Morrison, Hugh and {de Boer}, Gijs and Feingold, Graham and Harrington, Jerry and Shupe, Matthew D. and Sulia, Kara},
  year = 2012,
  journal = {Nature Geoscience},
  volume = {5},
  number = {1},
  pages = {11--17},
  publisher = {Nature Publishing Group},
  issn = {1752-0908},
  doi = {10.1038/ngeo1332},
  copyright = {2011 Springer Nature Limited}
}

@article{mulmenstadtFrequencyOccurrenceRain2015,
  title = {Frequency of Occurrence of Rain from Liquid-, Mixed-, and Ice-Phase Clouds Derived from {{A-Train}} Satellite Retrievals},
  author = {M{\"u}lmenst{\"a}dt, Johannes and Sourdeval, O. and Delano{\"e}, J. and Quaas, J.},
  year = 2015,
  journal = {Geophysical Research Letters},
  volume = {42},
  number = {15},
  pages = {6502--6509},
  issn = {1944-8007},
  doi = {10.1002/2015GL064604}
}

@article{nagelMsmhelperPythonPackage2023,
  title = {Msmhelper: {{A Python}} Package for {{Markov}} State Modeling of Protein Dynamics},
  shorttitle = {Msmhelper},
  author = {Nagel, Daniel and Stock, Gerhard},
  year = 2023,
  journal = {Journal of Open Source Software},
  volume = {8},
  number = {85},
  pages = {5339},
  issn = {2475-9066},
  doi = {10.21105/joss.05339}
}

@article{ovchinnikovIntercomparisonLargeeddySimulations2014,
  title = {Intercomparison of Large-Eddy Simulations of {{Arctic}} Mixed-Phase Clouds: {{Importance}} of Ice Size Distribution Assumptions},
  shorttitle = {Intercomparison of Large-Eddy Simulations of {{Arctic}} Mixed-Phase Clouds},
  author = {Ovchinnikov, Mikhail and Ackerman, Andrew S. and Avramov, Alexander and Cheng, Anning and Fan, Jiwen and Fridlind, Ann M. and Ghan, Steven and Harrington, Jerry and Hoose, Corinna and Korolev, Alexei and McFarquhar, Greg M. and Morrison, Hugh and Paukert, Marco and Savre, Julien and Shipway, Ben J. and Shupe, Matthew D. and Solomon, Amy and Sulia, Kara},
  year = 2014,
  journal = {Journal of Advances in Modeling Earth Systems},
  volume = {6},
  number = {1},
  pages = {223--248},
  issn = {1942-2466},
  doi = {10.1002/2013MS000282}
}

@article{pasquierNyAlesundAerosolCloud2022,
  title = {The {{Ny-\AA lesund Aerosol Cloud Experiment}} ({{NASCENT}}): {{Overview}} and {{First Results}}},
  shorttitle = {The {{Ny-\AA lesund Aerosol Cloud Experiment}} ({{NASCENT}})},
  author = {Pasquier, J. T. and David, R. O. and Freitas, G. and Gierens, R. and Gramlich, Y. and Haslett, S. and Li, G. and Sch{\"a}fer, B. and Siegel, K. and Wieder, J. and Adachi, K. and Belosi, F. and Carlsen, T. and Decesari, S. and Ebell, K. and Gilardoni, S. and {Gysel-Beer}, M. and Henneberger, J. and Inoue, J. and Kanji, Z. A. and Koike, M. and Kondo, Y. and Krejci, R. and Lohmann, U. and Maturilli, M. and Mazzolla, M. and Modini, R. and Mohr, C. and Motos, G. and Nenes, A. and Nicosia, A. and Ohata, S. and Paglione, M. and Park, S. and Pileci, R. E. and Ramelli, F. and Rinaldi, M. and Ritter, C. and Sato, K. and Storelvmo, T. and Tobo, Y. and Traversi, R. and Viola, A. and Zieger, P.},
  year = 2022,
  journal = {Bulletin of the American Meteorological Society},
  volume = {103},
  number = {11},
  pages = {E2533-E2558},
  publisher = {American Meteorological Society},
  issn = {0003-0007, 1520-0477},
  doi = {10.1175/BAMS-D-21-0034.1},
  chapter = {Bulletin of the American Meteorological Society}
}

@article{pernovComparisonSelectedSurface2024,
  title = {Comparison of Selected Surface Level {{ERA5}} Variables against In-Situ Observations in the Continental {{Arctic}}},
  author = {Pernov, Jakob Boyd and {Gros-Daillon}, Jules and Schmale, Julia},
  year = 2024,
  journal = {Quarterly Journal of the Royal Meteorological Society},
  volume = {150},
  number = {761},
  pages = {2123--2146},
  issn = {1477-870X},
  doi = {10.1002/qj.4700},
  copyright = {\copyright{} 2024 The Authors. Quarterly Journal of the Royal Meteorological Society published by John Wiley \& Sons Ltd on behalf of the Royal Meteorological Society.}
}

@article{pinskyAnalyticalInvestigationGlaciation2014,
  title = {Analytical {{Investigation}} of {{Glaciation Time}} in {{Mixed-Phase Adiabatic Cloud Volumes}}},
  author = {Pinsky, M. and Khain, A. and Korolev, A.},
  year = 2014,
  journal = {Journal of the Atmospheric Sciences},
  volume = {71},
  number = {11},
  pages = {4143--4157},
  publisher = {American Meteorological Society},
  issn = {0022-4928, 1520-0469},
  doi = {10.1175/JAS-D-13-0359.1},
  chapter = {Journal of the Atmospheric Sciences}
}

@article{pithanSelectStrengthsBiases2016,
  title = {Select Strengths and Biases of Models in Representing the {{Arctic}} Winter Boundary Layer over Sea Ice: The {{Larcform}} 1 Single Column Model Intercomparison},
  shorttitle = {Select Strengths and Biases of Models in Representing the {{Arctic}} Winter Boundary Layer over Sea Ice},
  author = {Pithan, Felix and Ackerman, Andrew and Angevine, Wayne M. and Hartung, Kerstin and Ickes, Luisa and Kelley, Maxwell and Medeiros, Brian and Sandu, Irina and Steeneveld, Gert-Jan and Sterk, H. a. M. and Svensson, Gunilla and Vaillancourt, Paul A. and Zadra, Ayrton},
  year = 2016,
  journal = {Journal of Advances in Modeling Earth Systems},
  volume = {8},
  number = {3},
  pages = {1345--1357},
  issn = {1942-2466},
  doi = {10.1002/2016MS000630},
  copyright = {\copyright{} 2016. The Authors.}
}

@article{raddatzDownwellingLongwaveRadiation2015,
  title = {Downwelling Longwave Radiation and Atmospheric Winter States in the Western Maritime {{Arctic}}},
  author = {Raddatz, R. L. and Papakyriakou, T. N. and Else, B. G. and Asplin, M. G. and Candlish, L. M. and Galley, R. J. and Barber, D. G.},
  year = 2015,
  journal = {International Journal of Climatology},
  volume = {35},
  number = {9},
  pages = {2339--2351},
  issn = {1097-0088},
  doi = {10.1002/joc.4149},
  copyright = {\copyright{} 2014 Royal Meteorological Society}
}

@article{rantanenArcticHasWarmed2022,
  title = {The {{Arctic}} Has Warmed Nearly Four Times Faster than the Globe since 1979},
  author = {Rantanen, Mika and Karpechko, Alexey Yu and Lipponen, Antti and Nordling, Kalle and Hyv{\"a}rinen, Otto and Ruosteenoja, Kimmo and Vihma, Timo and Laaksonen, Ari},
  year = 2022,
  journal = {Communications Earth \& Environment},
  volume = {3},
  number = {1},
  pages = {168},
  publisher = {Nature Publishing Group},
  issn = {2662-4435},
  doi = {10.1038/s43247-022-00498-3},
  copyright = {2022 The Author(s)}
}

@article{riechersStableStadialInterstadial2023,
  title = {Stable Stadial and Interstadial States of the Last Glacial's Climate Identified in a Combined Stable Water Isotope and Dust Record from {{Greenland}}},
  author = {Riechers, Keno and Rydin Gorj{\~a}o, Leonardo and Hassanibesheli, Forough and Lind, Pedro G. and Witthaut, Dirk and Boers, Niklas},
  year = 2023,
  journal = {Earth System Dynamics},
  volume = {14},
  number = {3},
  pages = {593--607},
  publisher = {Copernicus GmbH},
  issn = {2190-4979},
  doi = {10.5194/esd-14-593-2023}
}

@inbook{riskenFokkerPlanckEquationMethods1996,
  title = {The {{Fokker-Planck Equation}}: {{Methods}} of {{Solution}} and {{Applications}}},
  shorttitle = {The {{Fokker-Planck Equation}}},
  author = {Risken, Hannes},
  year = 1996,
  series = {Springer {{Series}} in {{Synergetics}}},
  volume = {18},
  publisher = {Springer},
  address = {Berlin, Heidelberg},
  doi = {10.1007/978-3-642-61544-3},
  copyright = {https://www.springernature.com/gp/researchers/text-and-data-mining},
  isbn = {978-3-540-61530-9 978-3-642-61544-3},
  chapter = {3}
}

@article{schemannSimulationMixedphaseClouds2020,
  title = {Simulation of Mixed-Phase Clouds with the {{ICON}} Large-Eddy Model in the Complex {{Arctic}} Environment around {{Ny-\AA lesund}}},
  author = {Schemann, Vera and Ebell, Kerstin},
  year = 2020,
  journal = {Atmospheric Chemistry and Physics},
  volume = {20},
  number = {1},
  pages = {475--485},
  publisher = {Copernicus GmbH},
  issn = {1680-7316},
  doi = {10.5194/acp-20-475-2020}
}

@article{schubertMarineStratocumulusConvection1979a,
  title = {Marine {{Stratocumulus Convection}}. Part {{II}}: {{Horizontally Inhomogeneous Solutions}}},
  shorttitle = {Marine {{Stratocumulus Convection}}. Part {{II}}},
  author = {Schubert, Wayne H. and Wakefield, Joseph S. and Steiner, Ellen J. and Cox, Stephen K.},
  year = 1979,
  journal = {Journal of the Atmospheric Sciences},
  volume = {36},
  number = {7},
  pages = {1308--1324},
  publisher = {American Meteorological Society},
  issn = {0022-4928, 1520-0469},
  doi = {10.1175/1520-0469(1979)036<1308:MSCPIH>2.0.CO;2},
  chapter = {Journal of the Atmospheric Sciences}
}

@article{schulteImprovingArcticMixedphase,
  title = {Towards Improving {{Arctic}} Mixed-Phase Cloud Representation in the {{ECMWF}} Model Using {{MOSAiC}} Observations},
  author = {Schulte, Luise and Forbes, Richard and Magnusson, Linus and Schemann, Vera and Day, Jonathan and Crewell, Susanne},
  journal = {Quarterly Journal of the Royal Meteorological Society},
  volume = {n/a},
  number = {n/a},
  pages = {e70205},
  issn = {1477-870X},
  doi = {10.1002/qj.70205},
  year = {2026},
  copyright = {\copyright{} 2026 European Centre for Medium Range Weather Forecasts and The Author(s). Quarterly Journal of the Royal Meteorological Society published by John Wiley \& Sons Ltd on behalf of Royal Meteorological Society.}
}

@article{shawUsingSatelliteObservations2022,
  title = {Using {{Satellite Observations}} to {{Evaluate Model Microphysical Representation}} of {{Arctic Mixed-Phase Clouds}}},
  author = {Shaw, J. and McGraw, Z. and Bruno, O. and Storelvmo, T. and Hofer, S.},
  year = 2022,
  journal = {Geophysical Research Letters},
  volume = {49},
  number = {3},
  pages = {e2021GL096191},
  issn = {1944-8007},
  doi = {10.1029/2021GL096191},
  copyright = {\copyright{} 2022. The Authors.}
}

@article{shupeOverviewMOSAiCExpedition2022,
  title = {Overview of the {{MOSAiC}} Expedition: {{Atmosphere}}},
  shorttitle = {Overview of the {{MOSAiC}} Expedition},
  author = {Shupe, Matthew D. and Rex, Markus and Blomquist, Byron and Persson, P. Ola G. and Schmale, Julia and Uttal, Taneil and Althausen, Dietrich and Angot, H{\'e}l{\`e}ne and Archer, Stephen and Bariteau, Ludovic and Beck, Ivo and Bilberry, John and Bucci, Silvia and Buck, Clifton and Boyer, Matt and Brasseur, Zo{\'e} and Brooks, Ian M. and Calmer, Radiance and Cassano, John and Castro, Vagner and Chu, David and Costa, David and Cox, Christopher J. and Creamean, Jessie and Crewell, Susanne and Dahlke, Sandro and Damm, Ellen and {de Boer}, Gijs and Deckelmann, Holger and Dethloff, Klaus and D{\"u}tsch, Marina and Ebell, Kerstin and Ehrlich, Andr{\'e} and Ellis, Jody and Engelmann, Ronny and Fong, Allison A. and Frey, Markus M. and Gallagher, Michael R. and Ganzeveld, Laurens and Gradinger, Rolf and Graeser, J{\"u}rgen and Greenamyer, Vernon and Griesche, Hannes and Griffiths, Steele and Hamilton, Jonathan and Heinemann, G{\"u}nther and Helmig, Detlev and Herber, Andreas and Heuz{\'e}, C{\'e}line and Hofer, Julian and Houchens, Todd and Howard, Dean and Inoue, Jun and Jacobi, Hans-Werner and Jaiser, Ralf and Jokinen, Tuija and Jourdan, Olivier and Jozef, Gina and King, Wessley and Kirchgaessner, Amelie and Klingebiel, Marcus and Krassovski, Misha and Krumpen, Thomas and Lampert, Astrid and Landing, William and Laurila, Tiia and Lawrence, Dale and Lonardi, Michael and Loose, Brice and L{\"u}pkes, Christof and Maahn, Maximilian and Macke, Andreas and Maslowski, Wieslaw and Marsay, Christopher and Maturilli, Marion and Mech, Mario and Morris, Sara and Moser, Manuel and Nicolaus, Marcel and Ortega, Paul and Osborn, Jackson and P{\"a}tzold, Falk and Perovich, Donald K. and Pet{\"a}j{\"a}, Tuukka and Pilz, Christian and Pirazzini, Roberta and Posman, Kevin and Powers, Heath and Pratt, Kerri A. and Preu{\ss}er, Andreas and Qu{\'e}l{\'e}ver, Lauriane and Radenz, Martin and Rabe, Benjamin and Rinke, Annette and Sachs, Torsten and Schulz, Alexander and Siebert, Holger and Silva, Tercio and Solomon, Amy and Sommerfeld, Anja and Spreen, Gunnar and Stephens, Mark and Stohl, Andreas and Svensson, Gunilla and Uin, Janek and Viegas, Juarez and Voigt, Christiane and {von der Gathen}, Peter and Wehner, Birgit and Welker, Jeffrey M. and Wendisch, Manfred and Werner, Martin and Xie, ZhouQing and Yue, Fange},
  year = 2022,
  journal = {Elementa: Science of the Anthropocene},
  volume = {10},
  number = {1},
  pages = {00060},
  issn = {2325-1026},
  doi = {10.1525/elementa.2021.00060}
}

@article{stevensModelIntercomparisonCCNlimited2018,
  title = {A Model Intercomparison of {{CCN-limited}} Tenuous Clouds in the High {{Arctic}}},
  author = {Stevens, Robin G. and Loewe, Katharina and Dearden, Christopher and Dimitrelos, Antonios and Possner, Anna and Eirund, Gesa K. and Raatikainen, Tomi and Hill, Adrian A. and Shipway, Benjamin J. and Wilkinson, Jonathan and Romakkaniemi, Sami and Tonttila, Juha and Laaksonen, Ari and Korhonen, Hannele and Connolly, Paul and Lohmann, Ulrike and Hoose, Corinna and Ekman, Annica M. L. and Carslaw, Ken S. and Field, Paul R.},
  year = 2018,
  journal = {Atmospheric Chemistry and Physics},
  volume = {18},
  number = {15},
  pages = {11041--11071},
  publisher = {Copernicus GmbH},
  issn = {1680-7316},
  doi = {10.5194/acp-18-11041-2018}
}

@article{stramlerSynopticallyDrivenArctic2011,
  title = {Synoptically {{Driven Arctic Winter States}}},
  author = {Stramler, Kirstie and Genio, Anthony D. Del and Rossow, William B.},
  year = 2011,
  journal = {Journal of Climate},
  volume = {24},
  number = {6},
  pages = {1747--1762},
  publisher = {American Meteorological Society},
  issn = {0894-8755, 1520-0442},
  doi = {10.1175/2010JCLI3817.1},
  chapter = {Journal of Climate}
}

@article{suraMultiplicativeNoiseNonGaussianity2005,
  title = {Multiplicative {{Noise}} and {{Non-Gaussianity}}: {{A Paradigm}} for {{Atmospheric Regimes}}?},
  shorttitle = {Multiplicative {{Noise}} and {{Non-Gaussianity}}},
  author = {Sura, Philip and Newman, Matthew and Penland, C{\'e}cile and Sardeshmukh, Prashant},
  year = 2005,
  journal = {Journal of the Atmospheric Sciences},
  volume = {62},
  number = {5},
  pages = {1391--1409},
  publisher = {American Meteorological Society},
  issn = {0022-4928, 1520-0469},
  doi = {10.1175/JAS3408.1},
  chapter = {Journal of the Atmospheric Sciences}
}

@inbook{rahimitabarAnalysisDataBasedReconstruction2019,
  title = {Analysis and {{Data-Based Reconstruction}} of {{Complex Nonlinear Dynamical Systems}}: {{Using}} the {{Methods}} of {{Stochastic Processes}}},
  shorttitle = {Analysis and {{Data-Based Reconstruction}} of {{Complex Nonlinear Dynamical Systems}}},
  author = {Rahimi Tabar, M. Reza},
  year = 2019,
  series = {Understanding {{Complex Systems}}},
  publisher = {Springer International Publishing},
  address = {Cham},
  doi = {10.1007/978-3-030-18472-8},
  copyright = {http://www.springer.com/tdm},
  isbn = {978-3-030-18471-1 978-3-030-18472-8},
  chapter = {20}
}

@book{Wegener1911,
      author = {A. Wegener},
      title = {{Thermodynamik der Atmosphäre}},
      year = {1911},
      publisher = {Barth},
      place = {Leipzig, Germany}
}

@article{woodsRoleMoistIntrusions2016,
  title = {The {{Role}} of {{Moist Intrusions}} in {{Winter Arctic Warming}} and {{Sea Ice Decline}}},
  author = {Woods, Cian and Caballero, Rodrigo},
  year = 2016,
  journal = {Journal of Climate},
  volume = {29},
  number = {12},
  pages = {4473--4485},
  publisher = {American Meteorological Society},
  issn = {0894-8755, 1520-0442},
  doi = {10.1175/JCLI-D-15-0773.1},
  chapter = {Journal of Climate}
}

@article{zuidemaArcticSpringtimeMixedPhase2005,
  title = {An {{Arctic Springtime Mixed-Phase Cloudy Boundary Layer Observed}} during {{SHEBA}}},
  author = {Zuidema, P. and Baker, B. and Han, Y. and Intrieri, J. and Key, J. and Lawson, P. and Matrosov, S. and Shupe, M. and Stone, R. and Uttal, T.},
  year = 2005,
  journal = {Journal of the Atmospheric Sciences},
  volume = {62},
  number = {1},
  pages = {160--176},
  publisher = {American Meteorological Society},
  issn = {0022-4928, 1520-0469},
  doi = {10.1175/JAS-3368.1},
  chapter = {Journal of the Atmospheric Sciences}
}

%%%%%%%%%%%%  Supplementary Figures  %%%%%%%%%%%%
% \clearpage

%%%%%%%%%%%% Supplementary Methods %%%%%%%%%%%%
% \footnotesize
% \section*{Supplementary material}
% \lipsum[101]

% \lipsum[102]

% \lipsum[103]

%%%%%%%%%%%%%%%%   End   %%%%%%%%%%%%%%%%
%\end{multicols}  % Method B for two-column formatting (doesn't play well with line numbers), comment out if using method A
\end{document}